\documentclass[useAMS,usenatbib]{mn2e}
\usepackage{amsmath}
\usepackage{amssymb}
\usepackage{graphicx}
\usepackage{lscape}
\usepackage{color}

\title {A new parameter space study of cosmological microlensing}
\author[G.~Vernardos et al.]
  {G.~Vernardos$^1$\thanks{gvernard@astro.swin.edu.au}, C.~J.~Fluke$^1$\\
  $^1$Centre for Astrophysics \& Supercomputing, Swinburne University of Technology, PO Box 218, Hawthorn, Victoria, 3122, Australia\\}
\pagerange{\pageref{firstpage}--\pageref{lastpage}} \pubyear{2012}

\begin{document}
\label{firstpage}
\maketitle

\begin{abstract}
Cosmological gravitational microlensing is a useful technique for understanding the structure of the inner parts of a quasar, especially the accretion disk and the central supermassive black hole.
So far, most of the cosmological microlensing studies have focused on single objects from $\sim$90 currently known lensed quasars.
However, present and planned all-sky surveys are expected to discover thousands of new lensed systems.
Using a graphics processing unit (GPU) accelerated ray-shooting code, we have generated 2550 magnification maps uniformly across the convergence ($\kappa$) and shear ($\gamma$) parameter space of interest to microlensing.
We examine the effect of random realizations of the microlens positions on map properties such as the magnification probability distribution (MPD).
It is shown that for most of the parameter space a single map is representative of an average behaviour.
All of the simulations have been carried out on the GPU-Supercomputer for Theoretical Astrophysics Research (gSTAR).
\end{abstract}

\begin{keywords}
gravitational lensing: micro -- accretion, accretion discs -- quasars: general
\end{keywords}

%$$$$$$$$$$$$$$$$$$$$$$$$$$$$$$$$$$$$$$$$$$$$$$$$$$$$$$$$$$$$$$$$$$$$$$$$$$$$$$$$$$$$$$$$$$$$$$$$$$$$$$$$$
\section{Introduction}
\label{sec:intro}
Understanding the properties of the supermassive black holes located at the centres of most galaxies \citep{Ferrarese2005}, is essential to our understanding of the evolution of the Universe.
The energy output observed in quasars, $\geq 10^{13}$ L$_{\odot}$ \citep{Peterson1997}, is believed to stem from the interaction of a central supermassive black hole with a surrounding accretion disc.
By studying accretion discs it is then possible to infer properties of the supermassive black holes, which in turn could constrain theories on their origin and evolution.

While plausible accretion disc models exist \citep[see ][for a review]{Abramowicz2013}, current observations are not sufficient to fully test them.
This is mainly because the angular diameter of the accretion disc at typical quasar distances, $\mathcal{O}(10^{-6})$ arcsec at redshift z$\sim$2, is orders of magnitude smaller than the resolution of our current best telescopes.
However, by detecting the imprints of gravitational microlensing over such cosmological scales we can extract useful information from the observations that can constrain accretion disc and, ultimately, black hole properties \citep{Rauch1991,Morgan2010}.

Light propagating from a distant quasar can be gravitationally lensed by a foreground galaxy resulting in the creation of luminous arcs, Einstein rings, multiple images, magnifications, etc.
In the case of an image of a background source being seen through the bulge of a foreground galaxy, individual stars that lie close to the line of sight can further deflect the light rays, acting as gravitational microlenses.
Their effect can be observed as uncorrelated variability in the light curve of a pair of images \citep{Paczynski1986}, or as anomalous flux ratios between the multiple images \citep{Witt1995,Schechter2002}.
Quasar microlensing was originally proposed by \cite{Chang1979}, and later detected by \cite{Irwin1989}.
For a recent review of the progress in this field see \cite{Schmidt2010}.

\begin{figure*}
\begin{center}
\includegraphics[width=\textwidth]{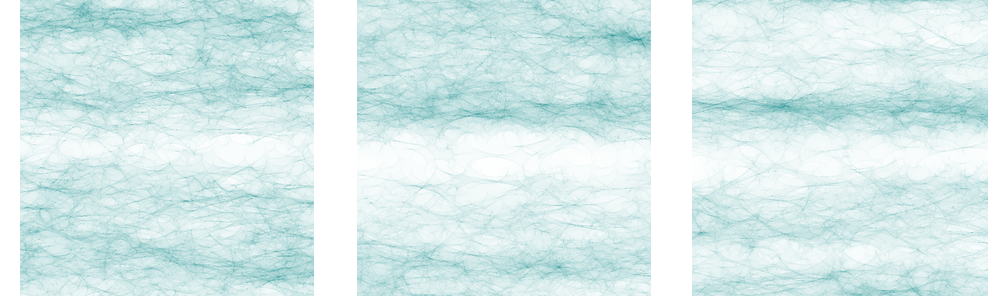}
\end{center}
\caption{Magnification maps produced for $\kappa,\gamma = (0.45,0.5)$, with a width of $24 R_{\rm Ein}$ and resolution of $4096^2$ pixels, resulting in 52752 individual microlenses distributed on the lens plane, for 3 sets of random microlens positions. High magnification areas are shown in blue. The 15 individual MPDs shown in Fig. \ref{fig:single} include the MPDs corresponding to these three maps. Although the map on the left looks different than the other two (larger area covered by caustics), it is the map in the middle that corresponds to the significantly different MPD (magenta line) of Fig. \ref{fig:single}.}
\label{fig:maps}
\end{figure*}

The collective effect of microlensing by individual stars in a foreground galaxy is most often investigated using a magnification, or caustic, map; a 2-dimensional pixellated map showing the magnification on a finite region of the source plane (Fig. \ref{fig:maps}).
The relative velocity of the observer, the source and the lens galaxy can take a background source from regions of low to high magnification on this map, inducing variability in the light curve.
The rate and amplitude of this variability is related to the size of the source with respect to the Einstein radius, $R_{\rm{Ein}}$, of the microlenses in the lensing galaxy; small sources will display larger and more rapid variations while extended ones will vary more smoothly \citep{Wambsganss1990b,Schmidt2010}.
Since quasar accretion discs are much smaller than the corresponding Einstein radii in all known systems \citep[$R_{\textrm{disc}}\lesssim 10^{15}$ cm compared to $R_{\textrm{Ein}} \simeq 10^{16}$ cm, see][]{Mosquera2011b}, they will be subject to microlensing essentially all the time.

Light-curve-analysis methods require continuous monitoring of a system for periods from a few months to several years \citep{Mosquera2011b}.
Such monitoring has been practical at some level for 20 to 30 systems \citep[for some examples see][and references therein]{Eigenbrod2007,Morgan2010}.
Alternatively, because gravitational microlensing is achromatic, single epoch multiwavelegth observations (snapshots) can estimate the accretion disc size in various wavelengths i.e. the temperature profile of the disc: $\lambda(T) \propto R^{\beta}$ \citep{Blaes2004,Abramowicz2013}.

The difficulties involved in both these methods of observing microlensed systems e.g. monitoring over large periods of time, coordinating observations at various wavelengths, measuring the lens and source redshifts, measuring time-delays etc., have kept the number of microlensed systems low; only $\sim$23 systems have been studied in detail to date \citep[][hereafter BF12]{Bate2012}.
Most studies have focused on single objects in the effort to constrain accretion disc properties, and only recently have collections of objects been considered \citep{Morgan2010,Blackburne2011,Mosquera2011a,Sluse2012}.
However, future planned all-sky surveys, like the Large Synoptic Survey Telescope \citep[LSST; ][]{LSST2009}, are expected to discover hundreds, or even thousands, of multiply imaged systems \citep{Oguri2010}.

Studying the parameter space of cosmological microlensing will be useful in understanding how the microlensing characteristics of a system can help constrain underlying accretion disc models.
Of particular interest is understanding how the properties of magnification maps, such as the magnification probability distribution (MPD), depend on the physical parameters convergence ($\kappa$) and shear ($\gamma$).
However, there are a number of additional parameters related to properties of the microlenses (mass, positions) and map characteristics (resolution, size, accuracy).
The imminent transition from single objects to statistically meaningful samples of microlensed quasars makes a parameter space approach to quasar microlensing timely (BF12).
%$$$$$$$$$$$$$$$$$$$$$$$$$$$$$$$$$$$$$$$$$$$$$$$$$$$$$$$$$$$$$$$$$$$$$$$$$$$$$$$$$$$$$$$$$$$$$$$$$$$$$$$$$

%$$$$$$$$$$$$$$$$$$$$$$$$$$$$$$$$$$$$$$$$$$$$$$$$$$$$$$$$$$$$$$$$$$$$$$$$$$$$$$$$$$$$$$$$$$$$$$$$$$$$$$$$$
\subsection{Explorations of parameter space}
A first exploratory study of microlensing parameter space was conducted by \citet[][hereafter W92]{Wambsganss1992}.
Although the resolution of the maps was low [e.g. to allow for an investigation of the power law behaviour of the MPD, $\mu^{-3}$, for high magnifications \citep{Schneider1987,Witt1990}], W92 provided some early evidence that the MPD is independent of the mass distribution of the microlenses.
\citet[][hereafter LI95]{Lewis1995} performed a similar parameter space exploration confirming this result.
A summary of the parameter values used in W92, LI95 and the present parameter space study (hereafter VF13) is shown in Table \ref{tab:1}.

Since W92 and LI95, no attempt to systematically explore parameter space has been made, mainly due to the lack of enough computational power to do so (W92).
Moreover, taking into account the low number of systems that exhibit microlensing, there has been little motivation in carrying out parameter space studies.
As a result, systems have been modelled individually.
However, the advent of supercomputers based on graphics processing units (GPUs) has enabled the systematic study of magnification maps across the entire microlensing parameter space (BF12).

The assumption of using a single map per combination of parameters, the usual approach in most microlensing studies, has not been examined in detail with respect to the random microlens positions.
Varying the microlens positions is an important test for the consistency of the existing microlensing techniques and for the robustness of the resulting accretion disc constraints.
Future high-resolution parameter space explorations, like the one proposed in BF12, should take into account any dependence of map properties on the microlens positions.

\begin{table}
\caption{Microlensing parameter space studies: Wambsganss (1992; W92), Lewis \& Irwin (1995; LI95) and the present one (VF13). $\boldsymbol{N}$ is the number of different values examined in each study for ($\kappa,\gamma$) combinations, $\kappa_{\textrm{s}}$, map size, mass of the microlenses (constant or distribution) and the set of random microlens positions (see Section \ref{sec:map-parameters}). The last row shows the total number of unique combinations of all the parameters in each study.}
\begin{center}
\begin{tabular}{|l|l|l|l|}
                                        & \textbf{W92} & \textbf{LI95} & \textbf{VF13} \\
\hline
\textbf{Method}                         & tree code    & contours      & direct        \\
$\boldsymbol{N_{\kappa,\gamma}}$        & 27           & 29            & 170           \\
$\boldsymbol{N_{\kappa_{\textrm{s}}}}$  & 16           & 1             & 1             \\
$\boldsymbol{N_{size}}$                 & 4            & 1             & 1             \\
$\boldsymbol{N_{mass}}$                 & 5            & 5             & 1             \\
$\boldsymbol{N_{positions}}$            & 1            & 1             & 15            \\
\hline
\textbf{Total}                          & 64           & 61            & 2550          \\
\end{tabular}
\end{center}
\label{tab:1}
\end{table}

In this work, we determine the number of magnification maps needed to fairly sample statistical properties, such as the MPD, for a given set of microlensing parameters.
We use the GPU-D \citep{Thompson2010} brute force ray-shooting code to uniformly cover the largest part of the $\kappa,\gamma$ parameter space.
We provide 170 $\kappa,\gamma$ combinations, producing 15 maps with different random microlens positions for each combination.
In Section 2 we introduce the direct inverse ray shooting technique and the microlensing parameter space.
In Section 3 we present our results, statistical properties of the MPD are extracted and an average behaviour per parameter space ($\kappa,\gamma$) pair is presented.
Discussion and conclusions follow in Sections 4 and 5.
%$$$$$$$$$$$$$$$$$$$$$$$$$$$$$$$$$$$$$$$$$$$$$$$$$$$$$$$$$$$$$$$$$$$$$$$$$$$$$$$$$$$$$$$$$$$$$$$$$$$$$$$$$

%$$$$$$$$$$$$$$$$$$$$$$$$$$$$$$$$$$$$$$$$$$$$$$$$$$$$$$$$$$$$$$$$$$$$$$$$$$$$$$$$$$$$$$$$$$$$$$$$$$$$$$$$$
\section{Method}
\label{sec:method}
At the heart of each microlensing technique lies the caustic magnification pattern created by compact objects and smooth matter in the lensing galaxy \citep{Kayser1986,Schneider1987}.
Once provided with a suitable map, one can study the background source (with a given size and/or profile) by either generating model light-curves or by calculating the MPD.
Key parameters for generating a magnification map are the convergence, $\kappa$, and shear, $\gamma$.
Convergence describes the focusing power of the microlenses and has two components, one from matter in compact objects (microlenses) $\kappa_{\textrm{*}}$, and another from smoothly distributed matter, $\kappa_{\textrm{s}}$.
Shear describes the external distortion applied to the images, which is dominated by the mass distribution of the lens galaxy.
The values of $\kappa$ and $\gamma$ arise from the macromodelling of the observed multiple images and the lensing galaxy \citep{Keeton2001}.
One can then use the lens equation to obtain the map:
\begin{equation}
\label{eq:lenseq}
\boldsymbol{y} = 
\begin{pmatrix}
1-\gamma & 0\\ 
0 & 1+\gamma 
\end{pmatrix}
\boldsymbol{x} -
\kappa _{\textrm{s}} \boldsymbol{x} - 
\sum_{i=1}^{N_{\textrm{l}}}m_{i} \frac{\left ( \boldsymbol{x - x _{i}} \right )}{\left | \boldsymbol{x -\boldsymbol{x _{i}}} \right | ^{2}} \, ,
\end{equation}
where $\boldsymbol{x}$ is in the image plane, $\boldsymbol{y}$ is in the source plane, $m_i$ and $\boldsymbol{x_i}$ are the mass and position of each microlens.
The number of microlenses, $N _{\textrm{l}}$, is related to $\kappa _{\textrm{*}}$ through:
\begin{equation}
\label{eq:Nl}
N_{\textrm{l}} = \frac{\kappa _{\textrm{*}} S}{\pi \langle M \rangle} \, ,
\end{equation}
for microlenses with mean mass $\langle M \rangle$, uniformly distributed over a surface S \citep[see][for an explanation of the choice of S]{Bate2010}.
The magnification is $\mu = 1 / det A$, with A being the Jacobian of the transformation from $\boldsymbol{y}$ to $\boldsymbol{x}$.

It was realized early on that analytic solutions of equation (\ref{eq:lenseq}) would exist only for a limited number of lens models \citep{Paczynski1986}.
In the case of quasar microlensing, $N_{\textrm{l}}$ can vary from tens to hundreds of thousands of microlenses, therefore, numerical methods had to be developed.
Most of these methods are based on the inverse ray-shooting technique introduced by \cite{Kayser1986} and \cite{Schneider1987}.
Light rays are propagated backwards from the observer, through the lens plane where they are deflected using equation (\ref{eq:lenseq}) and then mapped on to the source plane.
The source plane is divided into a grid of pixels.
By counting the number of rays reaching each pixel, $N_{ij}$, and comparing it to the number of rays that would have reached each pixel if there was no lensing, $N_{\rm{rays}}$, one obtains an estimate of the magnification:
\begin{equation}
\label{eq:mu}
\mu _{ij} = N _{ij} / N _{\rm{rays}}
\end{equation}
i.e. a pixelated magnification map.

A number of numerical methods to generate magnification maps exist in the literature.
\cite{Wambsganss1990a} used a hierarchical tree method to replace microlenses at a given distance from a light ray with a higher mass pseudo-lens, effectively reducing the final number of microlenses in the calculation.
\cite{Kochanek2004} allowed for the lens and source planes to be periodic and based his method on Fourier techniques (similar to $P^{3}M$ algorithms used in N-body problems), which reduced the number of microlenses as well.
\cite{Mediavilla2006,Mediavilla2011b} present an efficient method of inverse polygon mapping from the lens to the source plane which leads to the same accuracy as the previous approaches using far less rays \citep[same accuracy achieved in less than 1 per cent of the computational time,][]{Mediavilla2006}.
Finally, \cite{Thompson2010} made use of the high degree of parallelization inherent in the lens equation with a brute-force solution on a modern GPU.
\cite{Bate2010} compared compute times for this direct GPU approach and the more sophisticated tree-code approach and found it to be highly competitive (for a single CPU core), especially in regions of the parameter space where the number of microlenses is lower than $\sim 10^4$.
%$$$$$$$$$$$$$$$$$$$$$$$$$$$$$$$$$$$$$$$$$$$$$$$$$$$$$$$$$$$$$$$$$$$$$$$$$$$$$$$$$$$$$$$$$$$$$$$$$$$$$$$$$

%$$$$$$$$$$$$$$$$$$$$$$$$$$$$$$$$$$$$$$$$$$$$$$$$$$$$$$$$$$$$$$$$$$$$$$$$$$$$$$$$$$$$$$$$$$$$$$$$$$$$$$$$$
\subsection{Microlensing map parameters}
\label{sec:map-parameters}
The computation of the caustic structure in a microlensing magnification map depends on eight main parameters, which can be categorized into three groups:
\begin{description}
 \item Macro-model (external) parameters: $\boldsymbol{\kappa}$ and $\boldsymbol{\gamma}$ with physical interpretation as explained in Section \ref{sec:method}, and the smooth matter fraction, $\boldsymbol{s} = \kappa_{\rm s}/\kappa$, which describes the smooth matter component.
 \item Map characteristics and statistics: the map \textbf{size} (i.e. side length) should be large enough to enable simultaneous accretion disc and broad emission line region modelling. The map \textbf{resolution}\footnote{we are following the notation in our previous papers and use the imaging term of resolution: the total number of pixels in an image dimension.} should be high enough to resolve the inner parts of an accretion disc (see BF12 for a thorough discussion). The average number of rays per map pixel, $\boldsymbol{N_{\rm avg}}$, is a measure of the accuracy of a map and in general it is method dependent. For example, the polygon mapping technique \citep{Mediavilla2011b} uses more rays near the caustics and fewer away from them, where extra accuracy is not needed.
 \item Parameters of the microlenses: the \textbf{microlens mass}, $m_i$, appears explicitly in the lens equation (equation \ref{eq:lenseq}), while the mean microlens mass, $\langle M \rangle$, is involved in the calculation of the total number of microlenses (equation \ref{eq:Nl}). Microlensing studies can be performed once a mass function is chosen for stars in the lens galaxy. The \textbf{microlens positions}, $\boldsymbol{x_i}$ in equation (\ref{eq:lenseq}), cannot be measured by observations. Therefore, random realizations of the microlens positions have been used in generating magnification maps.
\end{description}
In most existing studies, only a single set of microlens positions is used under the assumption that different microlens configurations produce statistically equivalent maps.
In the present study we investigate the validity of this assumption by generating 15 maps with different sets of random microlens positions on the $\kappa,\gamma$ parameter space.
We now describe our choice of the remaining parameter values.

A complete treatment of the stellar mass function introduces several additional parameters (initial mass function, minimum mass cutoff, maximum mass etc).
Such a treatment is more practical in the case of single object studies; including it in a parameter space approach would result in additional computations.
Nevertheless, it is expected that the MPD will be relatively insensitive to the choice of the stellar mass function over most of $\kappa,\gamma$ parameter space (W92; LI95; \citealt{Wyithe2001}), except in some specific cases e.g. a highly magnified negative parity macroimage, as the one explored in \cite{Schechter2004} for $\kappa,\gamma = (0.55,0.55)$.
Therefore, we adopt the simplest treatment of the stellar mass function viz. a constant mass $m_i$ = 1 M$_{\odot}$ for all microlenses.

The total number of rays, for which we are calculating the deflections in equation (\ref{eq:lenseq}), is $\mathcal{O}(10^{10})$.
This results in an average number of rays per pixel, $N_{\rm avg}$, among the 2550 maps, equal to $302\pm24$.
\cite{Bate2010} and BF12 consider this number of rays to be sufficient for the statistical accuracy of the maps.
We performed a test of varying the map accuracy in selected regions of the $\kappa,\gamma$ parameter space, presented in Appendix A.
The results indicate that the suggestion of \cite{Bate2010} and BF12 is a reasonable choice.

We choose a size of 24 $R_{\rm{Ein}}$ for our maps and a resolution of 4096$^2$ pixels, following the strategy proposed in BF12.
This means that we can resolve sources down to 0.006 $R_{\rm{Ein}}$ in size.
If the size of a map is small compared to the typical size of a caustic ($\sim R_{\rm{Ein}}$), for fixed resolution, then a particular caustic structure may dominate the map and the corresponding MPD \citep[see fig. 1 in][ and panels 61-64 of fig. 1 in W92]{Wambsganss1990b}.
Therefore, varying the microlens positions on smaller maps is expected to have a larger effect.

In our present results we have assumed all matter to be in the form of compact objects i.e. $\kappa_{\textrm{s}} = 0$ for all maps.
For fixed $\kappa$, $N_{\textrm{l}}$ is maximized for $\kappa_{\textrm{s}}=0$; i.e. for any given $\kappa$, adding a contribution from smooth matter will reduce $\kappa_{\textrm{*}}$ which is proportional to the number of microlenses (equation \ref{eq:Nl}).
In this case, the statistical equivalence of maps with different microlens positions can be matched to the cases without smooth matter and a lower $\kappa$.
%$$$$$$$$$$$$$$$$$$$$$$$$$$$$$$$$$$$$$$$$$$$$$$$$$$$$$$$$$$$$$$$$$$$$$$$$$$$$$$$$$$$$$$$$$$$$$$$$$$$$$$$$$

\begin{figure}
\begin{center}
\includegraphics[width=0.47\textwidth]{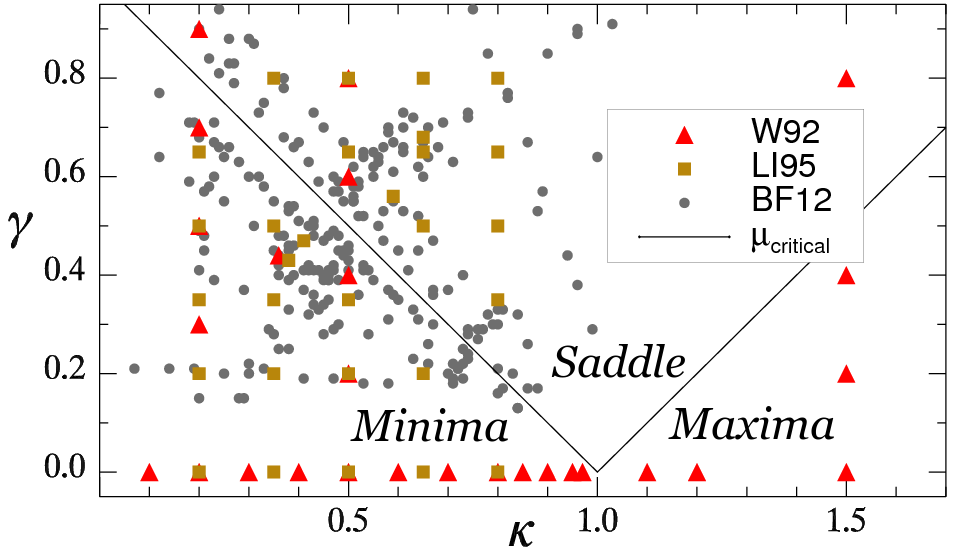}
\end{center}
\caption{Existing microlensing parameter space studies and macromodels. $\gamma$ is on the y-axis and $\kappa$ on the x-axis. Filled grey circles are from the compilation of BF12, filled red triangles from W92 and filled yellow squares from LI95. The critical line is shown in black and divides the $\kappa,\gamma$ parameter space in three areas: minima, saddle and maxima (see text).}
\label{fig:existing}
\end{figure}

%$$$$$$$$$$$$$$$$$$$$$$$$$$$$$$$$$$$$$$$$$$$$$$$$$$$$$$$$$$$$$$$$$$$$$$$$$$$$$$$$$$$$$$$$$$$$$$$$$$$$$$$$$
\subsection{The $\boldsymbol{\kappa},\boldsymbol{\gamma}$ parameter space}
The location in $\kappa,\gamma$-space of existing macromodels of multiply lensed systems and microlensing parameter space studies is presented in Fig. \ref{fig:existing}.
The grey circles show the results from the analysis of 85 macromodels of 23 known multiply imaged systems, as compiled and presented in BF12.
Each of these 85 models corresponds to two or four $\kappa,\gamma$ pairs, depending on whether the modelled system has a doubly or quadruply imaged background quasar.
Parameter space values investigated by W92 and LI95 are shown as red triangles and yellow squares respectively.

The macromagnification, $\mu_{\textrm{th}}$, for a given $\kappa,\gamma$ pair is:
\begin{equation}
\label{eq:muth}
\mu_{\textrm{th}} = \frac{1}{(1-\kappa)^2 - \gamma^2},
\end{equation}
with $\mu_{\textrm{th}} \rightarrow \infty$ when $\kappa=1 \pm \gamma$.
This is the critical line (black line in Fig. \ref{fig:existing}), which divides the parameter space into three distinct regions where macroimages can form: at the minima ($1-\kappa-\gamma>0$); saddle-points ($1-\kappa+\gamma<0$); and maxima ($1-\kappa-\gamma<0$) of the Fermat travel-time surface \citep{Blandford1986}.
Images found in these regions of parameter space exhibit different characteristics \citep{Schechter2002}, e.g. minima are always magnified ($ | \mu_{\textrm{th}} | > 1$), while saddle-points and maxima can be demagnified ($ | \mu_{\textrm{th}} | < 1$).
None of the 23 systems in the BF12 compilation has an observed maximum image, also called a central or odd image, as can be seen in Fig. \ref{fig:existing}.
Central images occur near the centre of the lensing galaxies, where the optical depth is high ($\kappa > 1$) and are thus subject to high demagnifications.
Although this makes their detection harder, central images may prove important for quasar microlensing \citep{Dobler2007}.

Fig. \ref{fig:existing} shows the present status of macromodelling of the 23 multiply imaged systems that are suitable for microlensing studies.
Once future synoptic all-sky surveys start to discover and monitor new multiply imaged quasars, new points will be added to this figure.
Moreover, alternative macromodels of existing or new systems can also increase the coverage of $\kappa,\gamma$ parameter space e.g. SDSS J0924+0219 is represented with 13 models \citep{Keeton2006,Morgan2006,Mediavilla2009}.
%$$$$$$$$$$$$$$$$$$$$$$$$$$$$$$$$$$$$$$$$$$$$$$$$$$$$$$$$$$$$$$$$$$$$$$$$$$$$$$$$$$$$$$$$$$$$$$$$$$$$$$$$$

%$$$$$$$$$$$$$$$$$$$$$$$$$$$$$$$$$$$$$$$$$$$$$$$$$$$$$$$$$$$$$$$$$$$$$$$$$$$$$$$$$$$$$$$$$$$$$$$$$$$$$$$$$
\section{Results}
\label{sec:results}
We have used the GPU-D code of \cite{Thompson2010} to generate 15 maps for each of 170 $\kappa$,$\gamma$ pairs: $0.05 \leq \kappa \leq 1.65$, $ 0.0 \leq \gamma \leq 0.9$, with $\Delta \kappa, \Delta \gamma = 0.1$.
Our choice of parameter space fully includes the critical line where the computations are more demanding.
As there have been no previous attempts at whole-parameter-space studies, we focused our effort on the more complex and challenging part of the parameter space along the critical line.
From the compilation of 85 lens macromodels presented in BF12, 95 per cent are found within our selected ranges of $\kappa,\gamma$.

All our simulations were carried out at gSTAR, the GPU Supercomputer for Theoretical Astrophysics Research, based at Swinburne University of Technology.
gSTAR made available to us a total of 50 compute nodes, each containing 2 NVIDIA Tesla C2070 GPUs which perform at greater than 1 Tflop/s (single precision).
Our dataset consists of a total of 2550 magnification maps at 4096$^2$ resolution, corresponding to 0.16 TB of data and 5100 GPU hours on gSTAR's C2070 cards.
%$$$$$$$$$$$$$$$$$$$$$$$$$$$$$$$$$$$$$$$$$$$$$$$$$$$$$$$$$$$$$$$$$$$$$$$$$$$$$$$$$$$$$$$$$$$$$$$$$$$$$$$$$

\begin{figure}
\begin{center}
\includegraphics[width=0.47\textwidth]{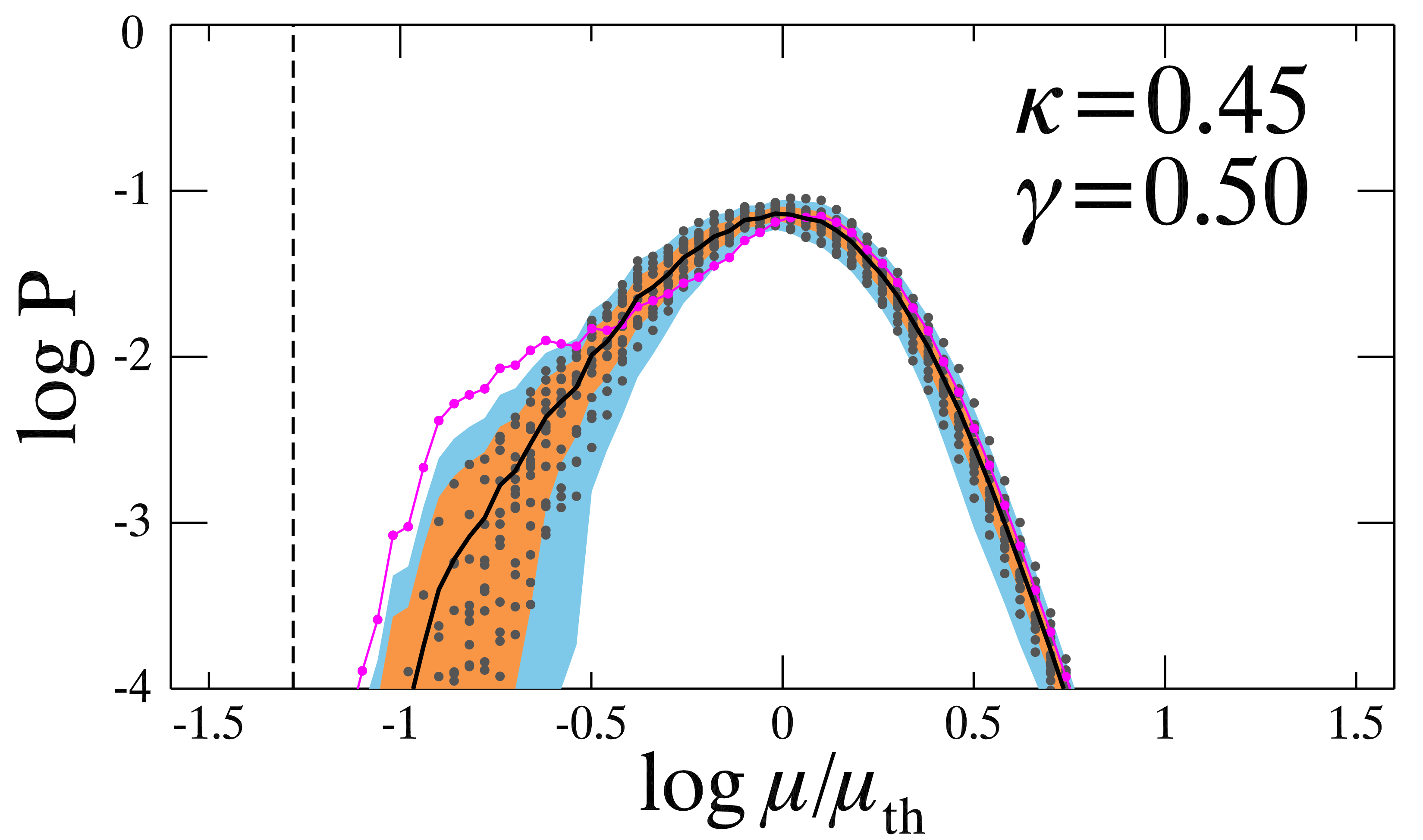}
\end{center}
\caption{A demonstration of the effect of different microlens positions on the MPD, for a single pair of $\kappa,\gamma$ values in parameter space. Dark grey dots represent 15 individual MPDs corresponding to maps with different microlens positions. The black line is the mean distribution, with the $\pm 1 \sigma$ and $\pm 2 \sigma$ limits shaded in orange and light blue. The dashed line is located at $-\mathrm{log}\,(\mu_{\mathrm{th}})$ and indicates the limit for no magnification i.e. $\mu=1$. We see that one of the distributions (magenta line) behaves differently than the rest for low magnifications. This MPD corresponds to the middle map in Fig. \ref{fig:maps}.}
\label{fig:single}
\end{figure}

%$$$$$$$$$$$$$$$$$$$$$$$$$$$$$$$$$$$$$$$$$$$$$$$$$$$$$$$$$$$$$$$$$$$$$$$$$$$$$$$$$$$$$$$$$$$$$$$$$$$$$$$$$
\subsection{Probability distributions}
One way of comparing different realizations of a magnification map is through the MPD.
By eye, the caustic structure in the three magnification maps shown in Fig. \ref{fig:maps} appears subtly different even though they all share the same macromodels.
E.g. the map on the left is almost completely covered by caustics, while the caustics on the map on the right appear clustered in horizontal bands.
The argument that different microlens positions lead to statistically equivalent maps means that although the maps in Fig. \ref{fig:maps} appear different, their MPDs should be similar.
In Fig. \ref{fig:single} we show the MPDs for 15 individual maps generated for $\kappa,\gamma = (0.45,0.5)$, where each map has different random microlens positions.
As we are covering only a limited range of $\mu$ (due to the finite number of pixels in a map), our MPDs are normalized as:
\begin{equation}
\label{eq:normsum}
\sum_{i=1}^{N_{\rm \mu}} \! P_i \, = 1 \, ,
\end{equation}
where $P_i$ is the probability to find a pixel of magnification $\mu_i$ in the map, and $N_{\rm \mu}$ is the total number of discrete $\mu_i$ values.
Each of the 15 MPDs is scaled by $\mu_{\rm th}$ and then binned in 100 logarithmic bins (grey points).
All 15 of the MPDs are similar for high magnifications, while one of them, shown in magenta, is different for low magnifications (log$\, \mu/\mu_{\rm th} \lesssim -0.4$).
This means that at least one of the 15 maps for the same $\kappa,\gamma$ may not be statistically equivalent to the rest.
Because the MPD is used in deriving accretion disc constraints (e.g. anomalous flux ratios), using a single map may lead to biased results.
In the following sections, we examine the statistical equivalence of the maps across the $\kappa,\gamma$ parameter space in more detail.

The thick black line in Fig. \ref{fig:single} is the average over the 15 individual distributions, and we refer to it in the following as the mean MPD or $\langle\textrm{MPD}\rangle$.
The orange shaded area shows the $\pm1 \sigma$ and the light blue shaded area the $\pm2 \sigma$ limits, where $\sigma$ is the standard deviation of the $\langle\textrm{MPD}\rangle$ in each bin.
The vertical dashed line indicates the limit for no magnification, i.e. $\mu=1$.

In Fig. \ref{fig:distrs} we present a tiling of the $\kappa,\gamma$ parameter space by the MPDs.
The range of the probability and the magnification in all diagrams is the same as in Fig. \ref{fig:single}.
The individual MPDs have been omitted for clarity.
The critical line on the $\kappa,\gamma$-space is shown as the thick black boundary between the tiles.
%$$$$$$$$$$$$$$$$$$$$$$$$$$$$$$$$$$$$$$$$$$$$$$$$$$$$$$$$$$$$$$$$$$$$$$$$$$$$$$$$$$$$$$$$$$$$$$$$$$$$$$$$$

\begin{figure*}
\begin{center}
%\vspace{21cm}
\includegraphics[height=21cm]{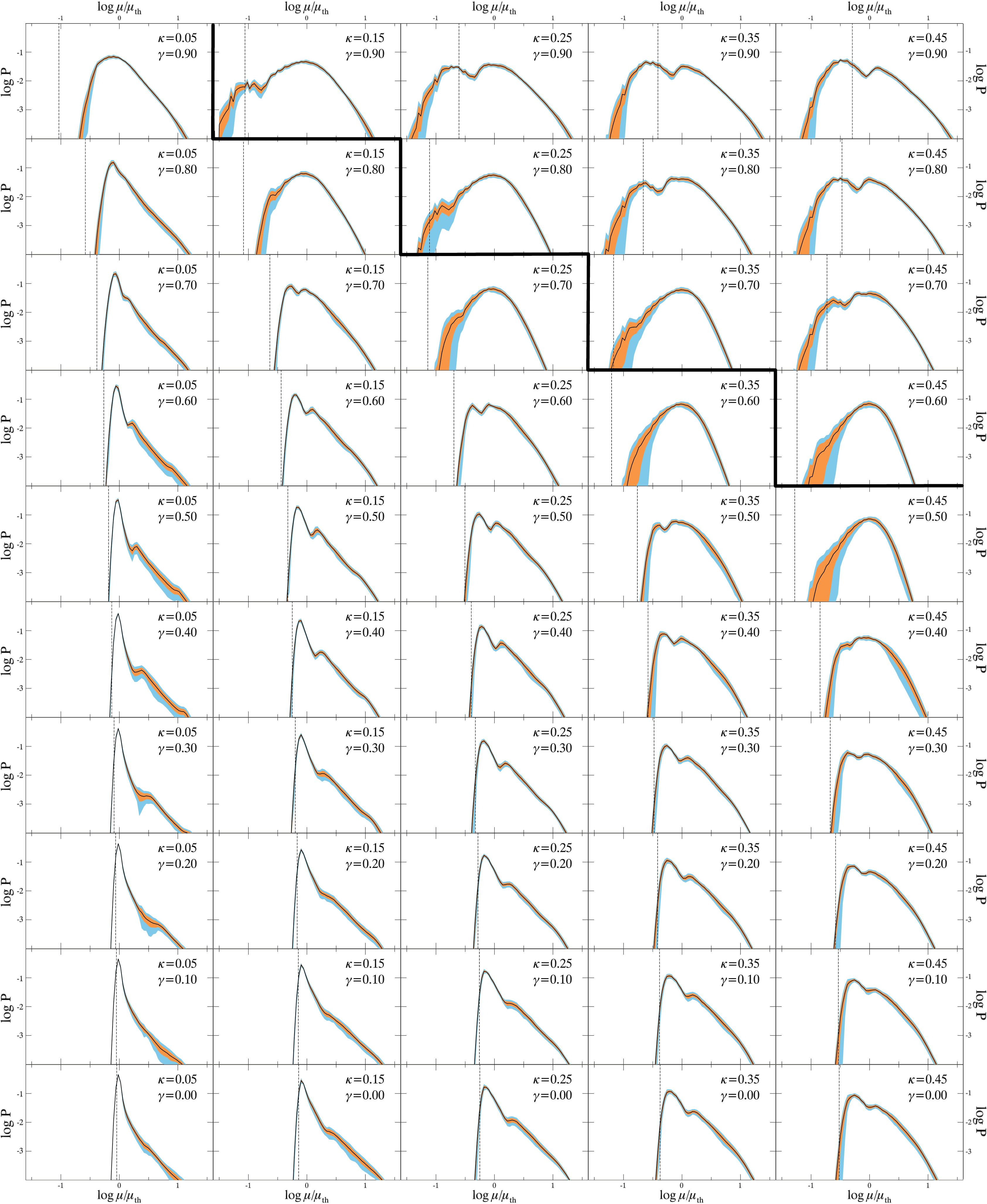}
\end{center}
\caption{Tiling of the $\kappa,\gamma$ parameter space by magnification probability distributions. In each tile we show the $\langle\textrm{MPD}\rangle$ (black line) and its $\pm 1$ (orange area) and $\pm 2$ (light blue area) $\sigma$ error, which were calculated from the underlying 15 magnification maps. Units and ranges are the same as in Fig. \ref{fig:single}. The dashed line is located at $-\mathrm{log}\,(\mu_{\mathrm{th}})$ and indicates the limit for no magnification i.e. $\mu=1$. The critical line on the $\kappa,\gamma$ parameter space is indicated by a thick black boundary between the tiles. 2550 magnification maps were used to generate this plot, corresponding to 5100 GPU hours. In this part, we see the distributions for $\kappa \leq 0.45$, with the panels below the critical line corresponding to the minima and the ones above to the saddle-point region.}
\label{fig:distrs}
\end{figure*}
\begin{figure*}
\begin{center}
%\vspace{21cm}
\includegraphics[height=21cm]{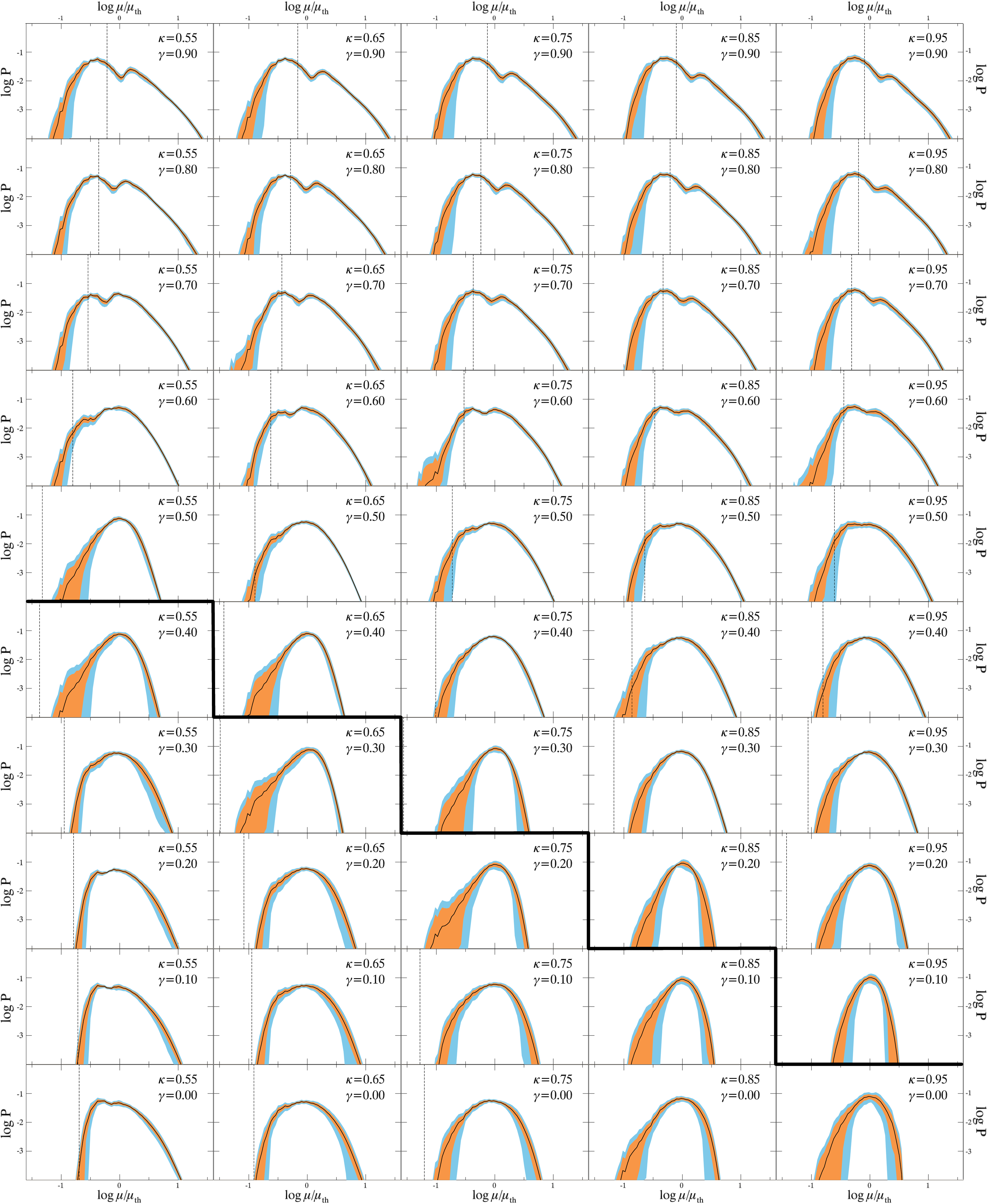}
\end{center}
\contcaption{The MPDs for $0.55 \leq \kappa \leq 0.95$. The panels below the critical line correspond to the minima and the ones above to the saddle-point region.}
\end{figure*}
\begin{figure*}
\begin{center}
%\vspace{21cm}
\includegraphics[height=21cm]{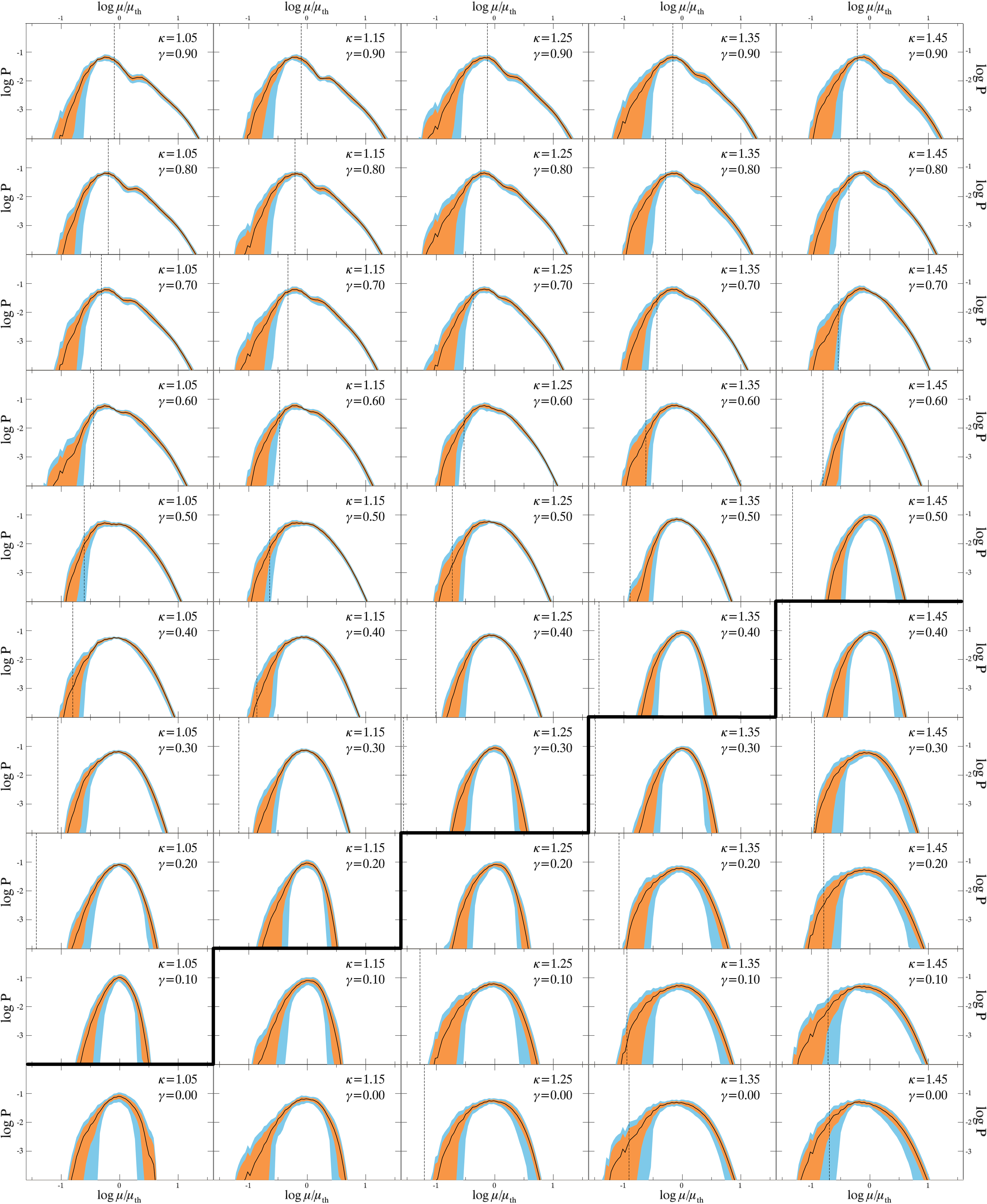}
\end{center}
\contcaption{The MPDs for $1.05 \leq \kappa \leq 1.45$. The panels below the critical line correspond to the maxima and the ones above to the saddle-point region.}
\end{figure*}
\begin{figure}
\begin{center}
%\vspace{21cm}
\includegraphics[height=21cm]{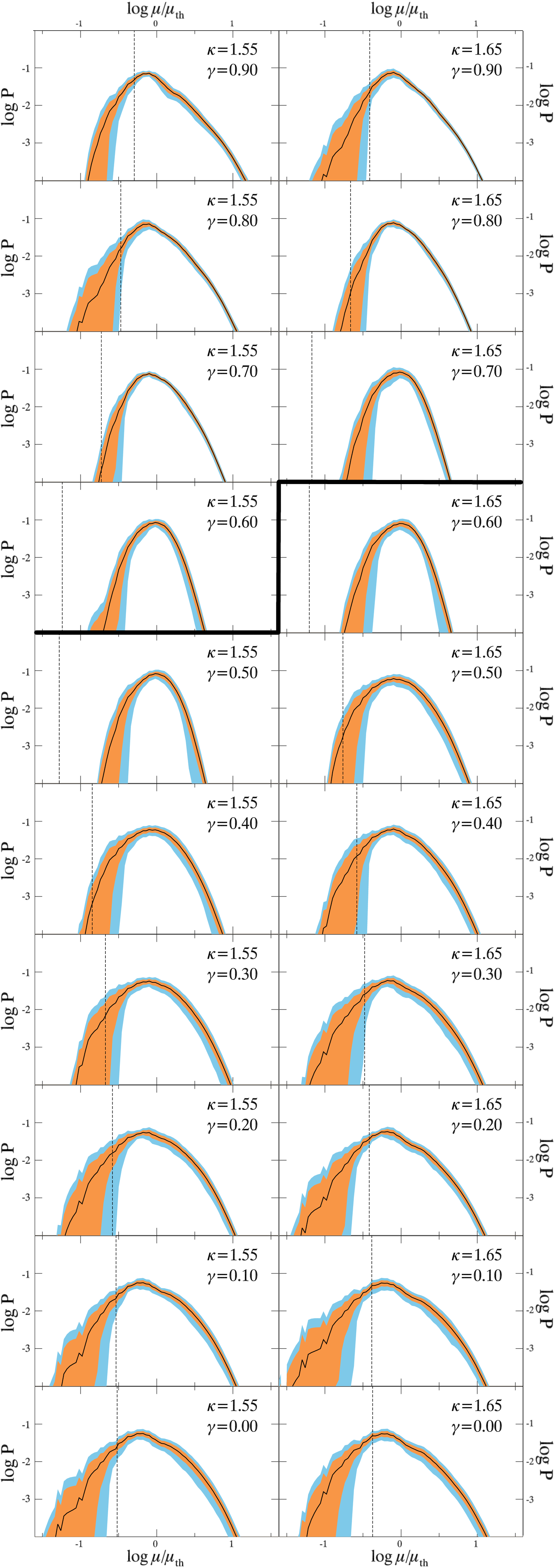}
\end{center}
\contcaption{The MPDs for $1.55 \leq \kappa \leq 1.65$. The panels below the critical line correspond to the maxima and the ones above to the saddle-point region.}
\end{figure}

%$$$$$$$$$$$$$$$$$$$$$$$$$$$$$$$$$$$$$$$$$$$$$$$$$$$$$$$$$$$$$$$$$$$$$$$$$$$$$$$$$$$$$$$$$$$$$$$$$$$$$$$$$
\subsection{Statistical properties}
In order to tackle the problem of statistical equivalence between the MPDs at given $\kappa,\gamma$ values for different microlens positions, we now examine their statistical properties.
We calculate the mean, the mode, the standard deviation, the median and the skewness for each of the 15 MPDs, for each $\kappa,\gamma$ pair.

The panels on the left-hand column of Fig. \ref{fig:stats} show the mean (noted as angle brackets enclosing the property under consideration) of each statistical property, while the right-hand column shows the corresponding standard deviation, s, among the 15 MPDs at each $\kappa,\gamma$ position.
In Figs \ref{fig:stats}(a), (b), (c) and (d) we show the mean of the mean, $\bar{\mu}$, the mode, Mo$\mu$, the standard deviation, $\sigma$, and the median, Md${\mu}$, of the MPDs, measured in units of the theoretical magnification $\mu_{\textrm{th}}$ (equation \ref{eq:muth}).
The corresponding standard deviation of these properties are shown in Figs \ref{fig:stats}(a'), (b'), (c') and (d') measured in units of $\mu_{\textrm{th}}$ as well.
Figs \ref{fig:stats}(e) and (e') show the logarithm of the mean and the standard deviation of the skewness, $g_1$, which is a dimensionless quantity.

In Fig. \ref{fig:stats}(a) we use a reference value of $\mu_{\textrm{th}} = 1$ (white on the color scale).
Greater values ($>\mu_{\rm{th}}$) are shown in red and lower ($<\mu_{\rm{th}}$) in blue.
We do the same for the mode, Fig. \ref{fig:stats}(b), but only the blue part of the colorbar is used because the mode is always found to be lower than $\mu_{\rm{th}} = 1$.
In Figs \ref{fig:stats}(c), (d) and (e) low values are shown in white and high values in red, but the range differs in each case.
Skewness was found to be always positive, with $\sim 7$ per cent of the cases having $g_1 > 10$, quite higher than the rest, therefore, we are showing the logarithm of the skewness and the corresponding standard deviation in Figs \ref{fig:stats}(e) and (e').
A consistent color scale (white to red) is used for all the standard deviation plots (panels on the right column of Fig. \ref{fig:stats}), starting always from zero and having a varying maximum value.
%$$$$$$$$$$$$$$$$$$$$$$$$$$$$$$$$$$$$$$$$$$$$$$$$$$$$$$$$$$$$$$$$$$$$$$$$$$$$$$$$$$$$$$$$$$$$$$$$$$$$$$$$$

%$$$$$$$$$$$$$$$$$$$$$$$$$$$$$$$$$$$$$$$$$$$$$$$$$$$$$$$$$$$$$$$$$$$$$$$$$$$$$$$$$$$$$$$$$$$$$$$$$$$$$$$$$
\subsection{Kolmogorov--Smirnov tests}
Along with calculating standard statistical properties of the MPDs, we also perform Kolmogorov--Smirnov (KS) tests between pairs of MPDs and between MPDs and the corresponding $\langle\textrm{MPD}\rangle$.

In the first case, the null hypothesis is that the MPDs are sampled from the same underlying distribution.
For each $\kappa,\gamma$ value we have 15 MPDs, which combine into 105 unique pairs.
We perform the KS test for each pair; if the resulting p-value is less than 0.05, then that particular pair of MPDs failed the test i.e. the MPDs are from different underlying distributions.
In Figs \ref{fig:KStests}(a), (b) and (c) we show the percentage of pairs that failed the KS test, $N_f$, across the $\kappa,\gamma$ parameter space, for MPDs binned in 100, 150 and 200 bins respectively.
The color scale ranges from 0 (white) to 100 (black) per cent.

In the second case, the null hypothesis is that each MPD and the corresponding $\langle\textrm{MPD}\rangle$ are originally from the same underlying distribution.
If the p-value of the test is less than 0.05, then the individual MPD failed the test with the $\langle\textrm{MPD}\rangle$.
The results for MPDs and $\langle\textrm{MPD}\rangle$ binned in 100, 150 and 200 bins are shown in Figs \ref{fig:KStests}(a'), (b') and (c').
The color code indicates the actual number of MPDs that failed the KS test with the mean, ranging from 0 (white) to 15 (black).

We perform the KS test between pairs of MPDs while also varying the side length of the map area from which the MPDs are calculated (with the number of bins fixed to 100).
From our 24 $R_{\rm{Ein}}$ maps, we are selecting central areas of widths equal to 20 $R_{\rm{Ein}}$, 16 $R_{\rm{Ein}}$ and 12 $R_{\rm{Ein}}$.
The number of pixels per $R_{\rm{Ein}}$ stays the same (170) in each case as in our original maps.
In Figs \ref{fig:KStests-width}(a), (b) and (c) we perform the KS test between pairs of MPDs for areas of our maps with widths equal to 20 $R_{\rm{Ein}}$, 16 $R_{\rm{Ein}}$ and 12 $R_{\rm{Ein}}$ respectively.
The color scale ranges from 0 (white) to 100 (black) per cent.
In Figs \ref{fig:KStests-width}(a'), (b') and (c') we perform the KS test between the MPDs of the smaller-sized maps and the $\langle\textrm{MPD}\rangle$ of the 24 $R_{\rm{Ein}}$ maps, which is our best approximation of the MPD for each $\kappa,\gamma$ value.
The color code indicates the actual number of the smaller-size map MPDs that failed the KS test with the $\langle\textrm{MPD}\rangle$ of the 24 $R_{\rm{Ein}}$ maps, and it ranges from 0 (white) to 15 (black).

Finally, we have chosen 5 pairs of $\kappa,\gamma$ values across the parameter space (the same which appear in Fig. \ref{fig:navg}) and generated maps with reduced width and resolution.
We have set the width and resolution equal to 20 $R_{\rm Ein}$ and 3413 pixels, 16 $R_{\rm Ein}$ and 2730 pixels and 12 $R_{\rm Ein}$ and 2048 pixels (the number of pixels per $R_{\rm Ein}$ has been kept the same as our original 24 $R_{\rm Ein}$ and 4096-pixel maps), and generated 15 maps with different microlens positions in each case (a total of 225 maps).
We find that the MPDs have the same form as the central regions of reduced width of our original 24 $R_{\rm Ein}$ and 4096-pixel maps, and display similar variation with respect to the different microlens positions.
Therefore, we can use our original maps across the parameter space to examine the effect of map width and microlens positions on the MPDs.
%$$$$$$$$$$$$$$$$$$$$$$$$$$$$$$$$$$$$$$$$$$$$$$$$$$$$$$$$$$$$$$$$$$$$$$$$$$$$$$$$$$$$$$$$$$$$$$$$$$$$$$$$$

\begin{figure*}
\begin{center}
%\vspace{21cm}
\includegraphics[width=\textwidth]{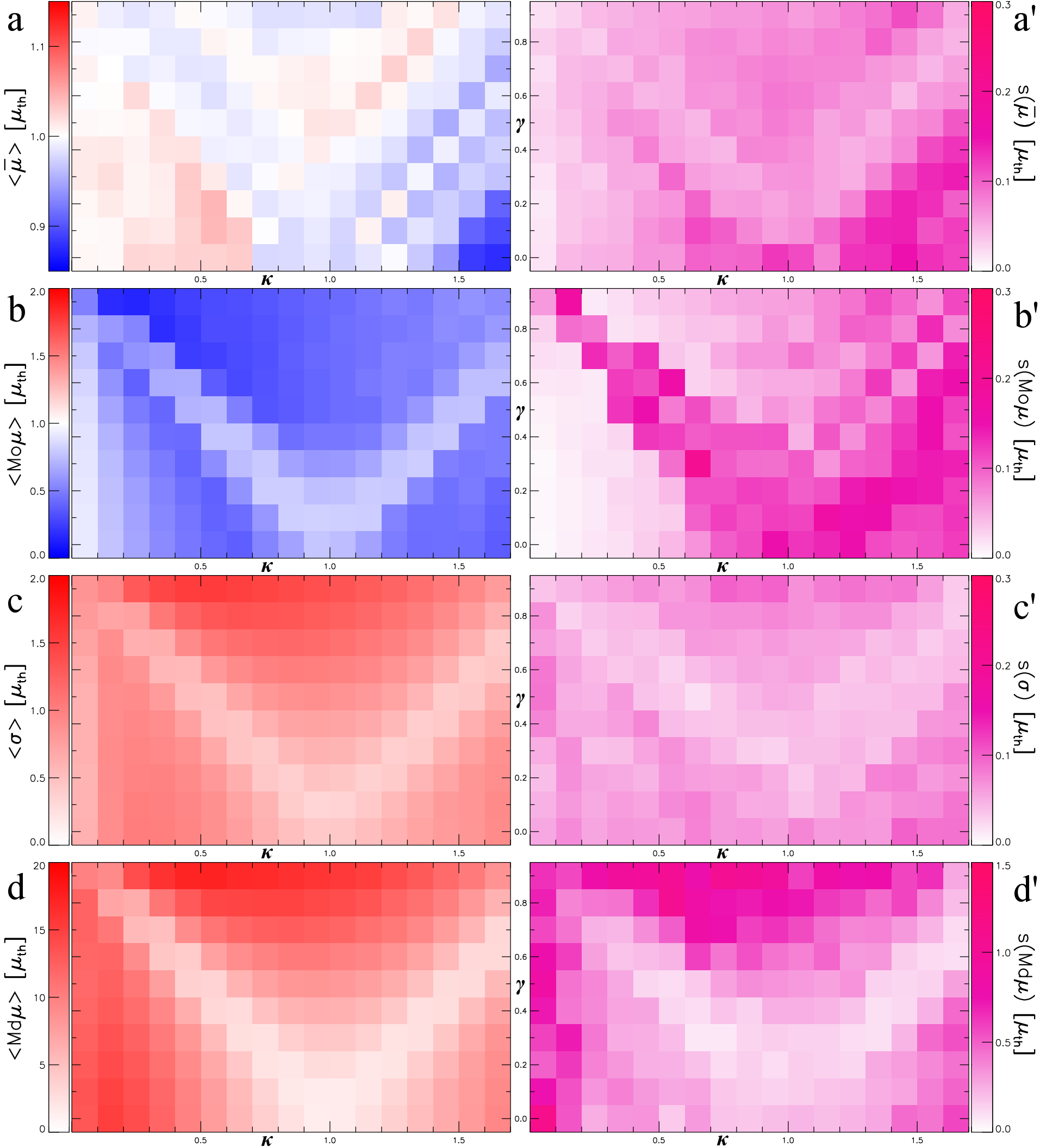}
\end{center}
\caption{Statistical properties of the MPDs across the $\kappa,\gamma$ parameter space, similar to Fig. \ref{fig:existing}.
Panels a,b,c,d show the mean of the mean, the mode, the standard deviation and the median across the parameter space and panels a',b',c',d' the standard deviation, of a sample of 15 cases per $\kappa,\gamma$ value. The mean and the mode are compared to $\mu_{\textrm{th}}$ which is always shown in white, with higher values shown in red and lower in blue. All the quantities are measured in units of $\mu_{\textrm{th}}$.}
\label{fig:stats}
\end{figure*}

\begin{figure*}
\begin{center}
%\vspace{5cm}
\includegraphics[width=\textwidth]{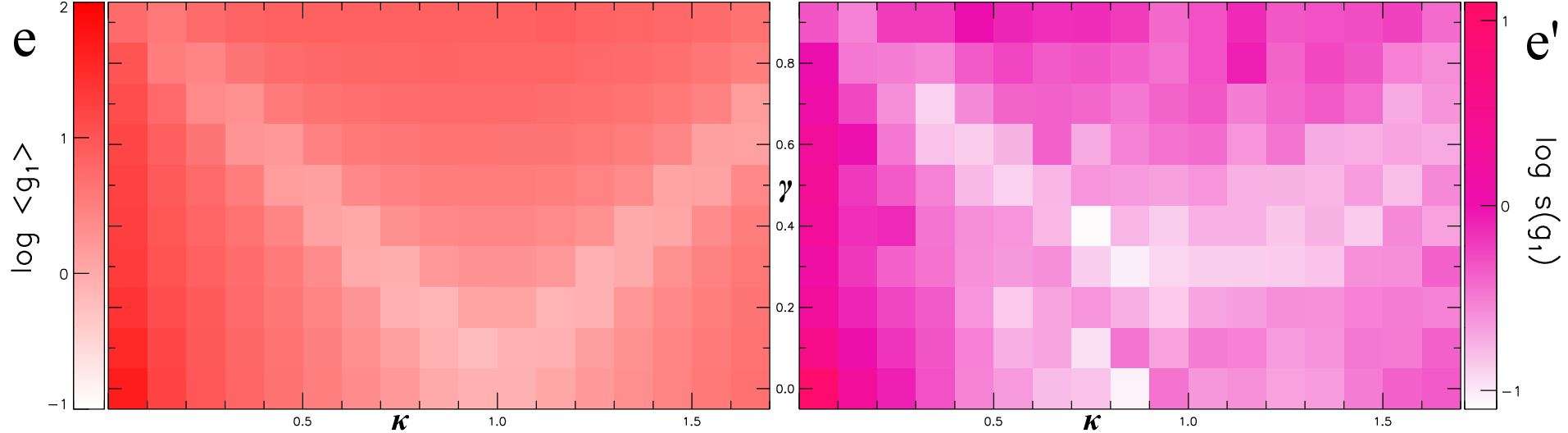}
\end{center}
\contcaption{Panel e shows the mean of the skewness of the MPDs and panel e' the standard deviation. Skewness is a dimensionless number and we found it to be always positive.}
\end{figure*}

\begin{figure*}
%\vspace{15cm}
\includegraphics[width=\textwidth]{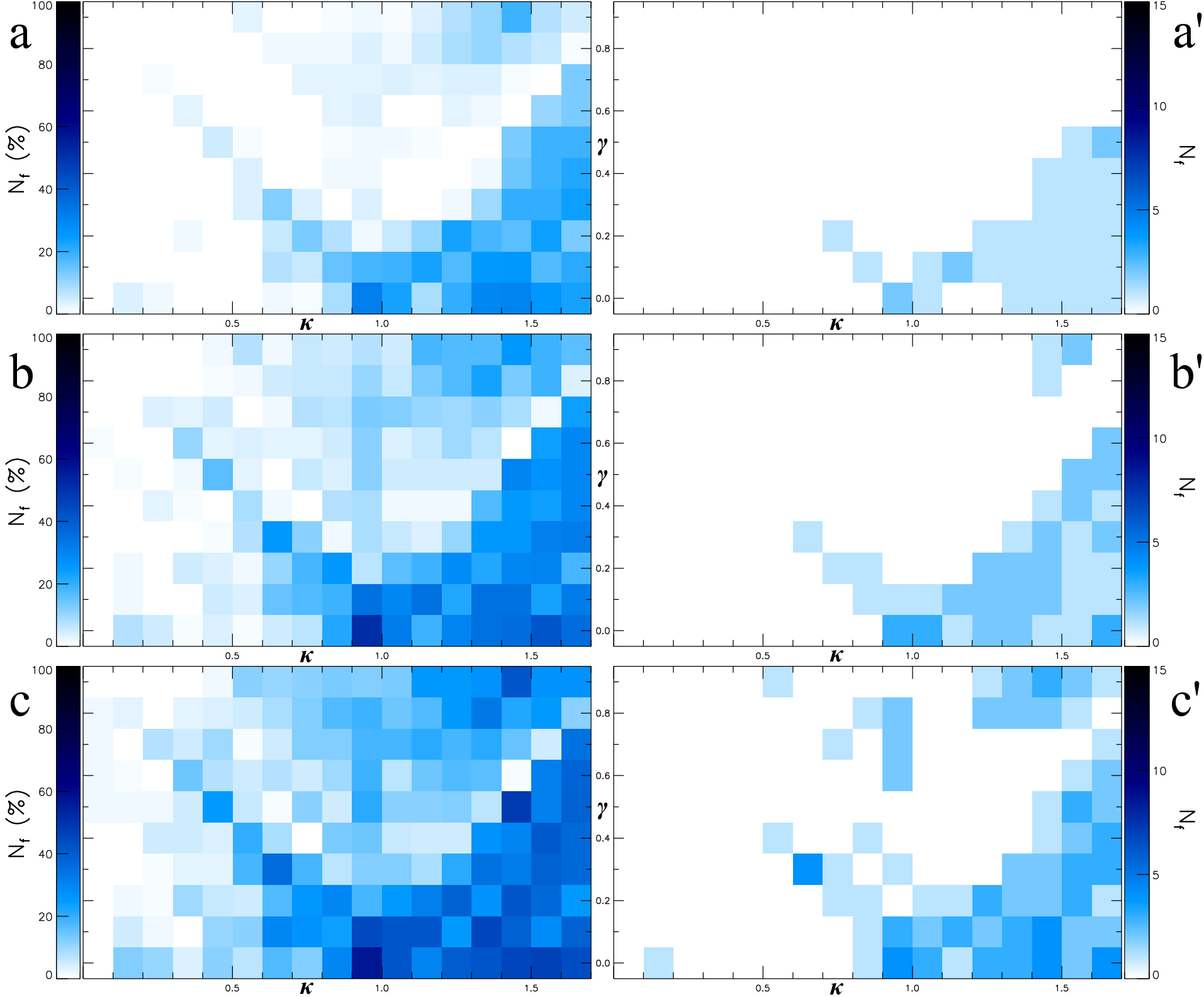}
\caption{KS tests in the $\kappa,\gamma$ parameter space for different MPD binning. In panels a, b and c we perform the KS test among pairs of MPDs setting the number of bins to 100, 150 and 200. The number of pairs that failed the test, $N_f$, is shown as a percentage of the total number of 105 pairs, for each $\kappa,\gamma$ value. In panels a',b' and c' we perform the KS test among the MPDs and the $\langle\textrm{MPD}\rangle$ setting the number of bins to 100, 150 and 200. For each $\kappa,\gamma$ value we show the number of MPDs, $N_f$, that failed the test with the $\langle\textrm{MPD}\rangle$.}
\label{fig:KStests}
\end{figure*}

\begin{figure*}
%\vspace{15cm}
\includegraphics[width=\textwidth]{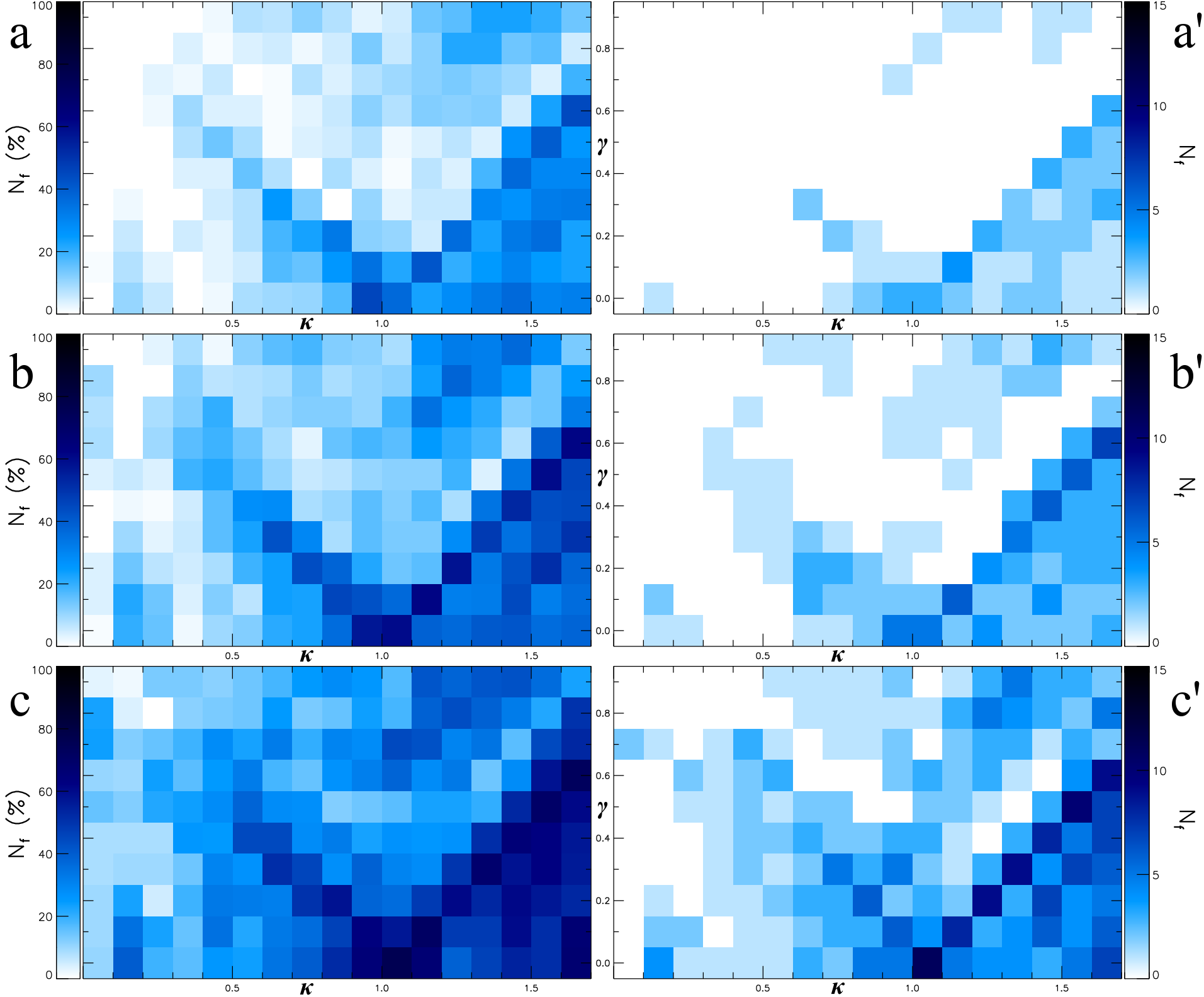}
\caption{KS tests in the $\kappa,\gamma$ parameter space for varying map width. In panels a, b and c we perform the KS test among pairs of MPDs extracted from a 20 $R_{\rm{Ein}}$, 16 $R_{\rm{Ein}}$ and 12 $R_{\rm{Ein}}$ wide area, located in the centre of our 24 $R_{\rm{Ein}}$ maps. The number of pairs that failed the test, $N_f$, is shown as a percentage of the total number of 105 pairs, for each $\kappa,\gamma$ value. In panels a',b' and c' we perform the KS test among the downsized MPDs and the $\langle\textrm{MPD}\rangle$ of the 24 $R_{\rm{Ein}}$ maps, which is our best approximation of the MPD for each $\kappa,\gamma$ value. At each $\kappa,\gamma$ value we show the number of MPDs, $N_f$, that failed the KS test.}
\label{fig:KStests-width}
\end{figure*}

\begin{table*}
\caption{Multiply imaged quasars for which the existing macromodels of the lensing galaxy (from the BF12 compilation) give $\kappa,\gamma$ pairs that fall into areas of the parameter space where the number of pairs of MPDs failing the KS test is above 10 per cent, or at least one MPD fails the test with the $\langle\textrm{MPD}\rangle$ (Fig. \ref{fig:KStests}). We identify the images of each system which fall into these areas and the particular macromodels as labelled by the corresponding authors. The images in parenthesis lie outside, but very close to such areas.}
\begin{center}
\begin{tabular}{|r|l|l|l|}
system         & image(s)    & model(s)              & reference \\
\hline
QJ0158-4325    & A           & f$_{M/L}$=0.1,0.2,0.3 & \cite{Morgan2008}        \\
MG0414+0534    & A1,A2,B     & b=1.5                 & \cite{Witt1995}          \\
SDSSJ0924+0219 & all         & f$_{M/L}$=0.0,0.1,0.2 & \cite{Morgan2006}        \\
HE1104-1805    & A           & f$_{M/L}$=0.1         & \cite{Morgan2008}        \\
PG1115+080     & A1,A2,C (B) & b=1.5                 & \cite{Witt1995}          \\
RXJ1131-1231   & B,C (A)     & f$_{M/L}$=0.1         & \cite{Dai2010}           \\
HE1413+117     & B,C (A)     & b=1.5                 & \cite{Witt1995}          \\
Q2237-0305     & all         & ``best fit''          & \cite{Wambsganss1994}    \\
\hline
\end{tabular}
\end{center}
\label{tab:2}
\end{table*}
%$$$$$$$$$$$$$$$$$$$$$$$$$$$$$$$$$$$$$$$$$$$$$$$$$$$$$$$$$$$$$$$$$$$$$$$$$$$$$$$$$$$$$$$$$$$$$$$$$$$$$$$$$

%$$$$$$$$$$$$$$$$$$$$$$$$$$$$$$$$$$$$$$$$$$$$$$$$$$$$$$$$$$$$$$$$$$$$$$$$$$$$$$$$$$$$$$$$$$$$$$$$$$$$$$$$$
\section{Discussion}
\label{sec:discussion}
We produced microlensing magnification maps uniformly covering most of the $\kappa,\gamma$ parameter space of interest, using 15 different sets of random microlens positions at each $\kappa,\gamma$ value.
The effect of changing the random microlens positions on the MPD varies across the $\kappa,\gamma$ parameter space and depends on the map width.
This was identified in W92, who gave an example of four different map widths for the same $\kappa,\gamma$ and microlens positions (W92, cases 61-64 of table 1).
However, using our systematic parameter space approach we can estimate the number/width required for magnification maps to show an average behavior.

%$$$$$$$$$$$$$$$$$$$$$$$$$$$$$$$$$$$$$$$$$$$$$$$$$$$$$$$$$$$$$$$$$$$$$$$$$$$$$$$$$$$$$$$$$$$$$$$$$$$$$$$$$
\subsection{Statistical properties of the maps and the mean distribution}
Having 15 maps per $\kappa,\gamma$ point makes it possible to examine average statistical properties of the MPDs across the parameter space.
We calculate the mean, the mode, the standard deviation, the median and the skewness of the MPDs per $\kappa,\gamma$ point, as shown in Fig. \ref{fig:stats}.
We also calculate the $\langle\textrm{MPD}\rangle$ among the 15 cases per $\kappa,\gamma$ value and use it as an average MPD at each point of parameter space.

The mean magnification of the MPDs, $\bar{\mu}$ in Figs \ref{fig:stats}(a) and (a'), stays almost always within 5 per cent of the theoretical value $\mu_{\textrm{th}}$ (equation \ref{eq:muth}), in agreement with what was found in W92, which is explained by the finite size of the maps (24 $R_{\rm Ein}$).
However, $\bar{\mu}$ has a systematic behaviour when compared to $\mu_{\textrm{th}}$ in the parameter space: $\bar{\mu} \geq \mu_{\textrm{th}}$ in the minima, $\bar{\mu} \leq \mu_{\textrm{th}}$ in the maxima, while in the saddle-point region the situation is unclear.
This behaviour agrees with the expectation of minima being always magnified ($ | \mu_{\textrm{th}} | > 1$), while saddle-points and maxima can be demagnified ($ | \mu_{\textrm{th}} | < 1$) as well \citep{Schechter2002}.

Throughout the parameter space, the $\langle\textrm{MPD}\rangle$ appears to be dominated by two peaks and matches those produced by W92 and LI95 for the cases they studied.
The highest peak corresponds to the mode of the MPDs [Figs \ref{fig:stats}(b) and (b')], the maximum probability, and it is located at $\approx \mu_{\rm th}$ as expected.
The secondary peak has been identified previously in the works of \cite{Rauch1992}, W92 and LI95, and its origin has been attributed to the creation of very bright pairs of micro-images at caustics.
It is predicted that higher order pairs would appear at even higher magnification, effectively creating additional weaker peaks, which, however, require higher resolution to be seen.
Our $4096^2$ resolution is sufficient to confirm this by observing a third peak in at least six cases: $\kappa=0.05$ with $\gamma=0.4, 0.5, 0.6$ and $\kappa=0.15$ with $\gamma=0.3, 0.4, 0.5$ (Fig. \ref{fig:distrs}).

There seems to be an interplay between the two peaks of the $\langle\textrm{MPD}\rangle$ across the parameter space.
In the minima region we have a relatively narrow peak located just below $\mu_{\textrm{th}}$, together with a second wider peak that appears at higher magnification.
This explains the location of the mode of the MPDs below $\mu_{\textrm{th}}$, and the high values of the standard deviation, the median and the skewness (Fig. \ref{fig:stats}); positive skewness means assymetric MPDs with tails expanding into high magnifications.
As we approach the critical line, the secondary peak approaches $\mu_{\textrm{th}}$ and gradually dominates as the primary peak vanishes.
Close to the critical line, only one peak is present and located at $\mu \approx \mu_{\textrm{th}}$.
The mode is closest to $\mu_{\textrm{th}}$, the standard deviation, the median and the skewness get their minimum values; the MPDs become narrow and symmetric.
Crossing into the saddle-point region, for $\gamma < 0.5$ only one peak remains while for $\gamma > 0.5$ we have two prominent peaks, one in low and one in high magnification, which cause the values of the standard deviation and the median to increase.
As we approach again the critical line at higher optical depth, the low magnification peak approaches $\mu_{\textrm{th}}$ and gradually dominates, while the high magnification peak vanishes.
Reaching the critical line at $\kappa > 1$ from the saddle-point region, there is again only one peak located at $\mu \approx \mu_{\textrm{th}}$ and the MPDs are again narrow and symmetric.
This peak persists as we go from near the critical line into the maxima region, but becomes wider and reaches into demagnifications the further we go.
Using the results described above we can conclude that:
\begin{enumerate}
 \item Along the critical line, the MPDs are centered around $\mu \approx \mu_{\textrm{th}}$, extend in a narrow area around it, $0.2 < \mu/\mu_{\textrm{th}} < 5$, and appear symmetric (in log space).
 \item Away from the critical line, the MPDs generally become wider and assymetric, mostly due to the appearance of a high magnification tail and/or a secondary peak.
\end{enumerate}
There seems to be a smooth transition in the shape of the $\langle\textrm{MPD}\rangle$ across the parameter space.
This could mean that a fit of an analytical formula to the mean MPD could be possible.
In other words, an empirical MPD at \textbf{any} given point in $\kappa,\gamma$ parameter space could be constructed.
Moreover, posessing a $\langle\textrm{MPD}\rangle$ makes it possible to calculate the slope at high $\mu$, similar to W92, but in all points in $\kappa,\gamma$ parameter space and compare it to theoretical predictions \citep{Schneider1987,Witt1990}.
Performing these tasks is beyond the scope of this study and we will deal with them in the future.

A narrow strip of parameter space for $\kappa \approx 0.05$ is worthy of a special mention.
The standard deviation appears significantly lower in this strip than its adjacent neighbours at $\kappa \approx 0.1$.
This does not happen for the median and the skewness, which get their maximum values in this strip.
The explanation is that the MPDs located at $\kappa \approx 0.05$ are very narrow and have an extended high magnification tail.
This tail is dominated by the appearance of a secondary peak in the MPD.
%$$$$$$$$$$$$$$$$$$$$$$$$$$$$$$$$$$$$$$$$$$$$$$$$$$$$$$$$$$$$$$$$$$$$$$$$$$$$$$$$$$$$$$$$$$$$$$$$$$$$$$$$$

%$$$$$$$$$$$$$$$$$$$$$$$$$$$$$$$$$$$$$$$$$$$$$$$$$$$$$$$$$$$$$$$$$$$$$$$$$$$$$$$$$$$$$$$$$$$$$$$$$$$$$$$$$
\subsection{Interpreting the KS tests}
A known systematic of the KS test is related to the choice of the nubmer of bins for the distributions being compared.
A large number of bins picks up the fine features of each distribution, leading frequently to a p-value lower than 5 per cent and the test failing.
On the other hand, a small number of bins smooths out the differences among the distributions.
Because the binning of the MPDs is arbitrary, we performed the KS test for 3 different numbers of bins, namely 100, 150 and 200.
The anticipated behaviour of more MPDs failing the test as the number of bins grows is seen in Figs \ref{fig:KStests}(a), (b) and (c).
A number of 100 bins seems to describe the MPD in enough detail for the purposes of accretion disc modelling.

In Fig. \ref{fig:KStests}(a), we show the percentage of all possible MPD pairs at each $\kappa,\gamma$ value that failed the KS test, $N_f$.
$N_f$ is zero or lower than 5-10 per cent for most of the minima and the saddle point region and increases to around 20-30 per cent in the maxima region.
In an extreme case, where particular sets of random microlens positions lead to significantly different MPDs, the presence of one such map in our sample of 15 would result in 14 failed KS tests ($N_f \approx 13$ per cent) at the given $\kappa,\gamma$ value.
In practice, it is possible to have a single MPD for a given $\kappa,\gamma$ value that behaves different to the rest (e.g. magenta line in Fig. \ref{fig:single}), but it will fail the test with some and not all of the remaining 14 MPDs.
This is what we generally observe in our results, with $N_f$ around 10-20 per cent being mostly due to 2 or 3 particular MPDs.
However, all the 15 maps, and the corresponding MPDs, represent equally valid choices of random microlens positions, and the fact that some of them appear different to the rest is most likely due to random clustering of microlenses on the image plane.
Therefore, in the case of a pair of MPDs failing the KS test, we do not have any criterion yet to distinguish which of the two MPDs should be considered different; they are both equally ``correct''.
A way out is to perform the KS test between each of the 15 MPDs and the computed $\langle\textrm{MPD}\rangle$ for each $\kappa,\gamma$ value.
In this case, the $\langle\textrm{MPD}\rangle$ is more ``correct'' than a simple MPD; it represents an average behaviour of 15 MPDs.
If the KS test fails in this case, it is because that particular MPD is different to all the remaining 14 that are represented by their $\langle\textrm{MPD}\rangle$.
The results of such a test are shown in Figs \ref{fig:KStests}(a'), (b') and (c').
In the maxima region we have at least 1 out of 15 maps per $\kappa,\gamma$ value, whose MPD is different to the rest.
This also occurs on the critical line and for $0.7 < \kappa < 1.0$.
Increasing the number of bins leads to more MPDs failing the test with the $\langle\textrm{MPD}\rangle$ in larger areas of parameter space, similarly to the results of the KS test among pairs of distributions.

The combined effect of random microlens positions and map width can be seen in Fig. \ref{fig:KStests-width} throughout the $\kappa,\gamma$ parameter space.
For small map widths, $\sim$12 $R_{\rm Ein}$, the MPDs significantly vary for different random microlens positions.
As the map width is increased, the MPDs are converging.
Our maps have a 24 $R_{\rm Ein}$ side length, with atypical maps appearing mainly in the maxima area, at a rate of 1 or 2 out of 15.
Therefore, to achieve an average behaviour in the maxima area more than one map is required, or alternatively, maps wider than 24 $R_{\rm Ein}$ should be generated.

It should be noted here, that the regions of parameter space where there is a higher probability to have an atypical map correlate with regions of the highest number of microlenses, $N_l$ (equation \ref{eq:Nl}); in general, $N_l$ increases with $\kappa$ and gets its highest values for $\kappa,\gamma$ combinations, where $\mu_{\rm th}$ in equation \ref{eq:muth} approaches infinity i.e. along the critical line.
The contrary behaviour was anticipated: maps with low numbers of microlenses should be more prone to lens position systematics (e.g. random clustering in the lens plane) while for higher numbers ($\sim 10^9$) the purely smooth matter limit should be reached \citep{Garsden2012}.
The highest number of microlenses in our maps does not exceed $2.5 \times 10^5$ (for $\kappa = 1.05$ and $\gamma = 0.0$), therefore, further investigation is warranted to determine the cause for atypical maps appearing in this intermediate region.
Unfortunately, this is exactly where the computations are the most demanding.
%$$$$$$$$$$$$$$$$$$$$$$$$$$$$$$$$$$$$$$$$$$$$$$$$$$$$$$$$$$$$$$$$$$$$$$$$$$$$$$$$$$$$$$$$$$$$$$$$$$$$$$$$$

%$$$$$$$$$$$$$$$$$$$$$$$$$$$$$$$$$$$$$$$$$$$$$$$$$$$$$$$$$$$$$$$$$$$$$$$$$$$$$$$$$$$$$$$$$$$$$$$$$$$$$$$$$
\section{Conclusions}
We have performed a uniform systematic study of microlensing parameter space in preparation for future upcoming all-sky surveys.
We have shown that at 24 $R_{\rm Ein}$ map width, a single map should display an average behaviour, except in the maxima region where 2 or 3 maps should be considered.
For maps smaller than 24 $R_{\rm Ein}$, the effect of the microlens positions is stronger, leading to a more frequent occurence of atypical maps, even in the minima and saddle-point regions.
For maps wider than 24 $R_{\rm Ein}$, the corresponding MPDs are expected to display an average behaviour independently of the microlens positions.

We have introduced the $\langle\textrm{MPD}\rangle$ in Fig. \ref{fig:distrs} as a representative MPD of each point in the $\kappa,\gamma$ parameter space.
The MPD of any magnification map, independently of the method used to generate it, can be compared to the $\langle\textrm{MPD}\rangle$ shown in Fig. \ref{fig:distrs}.

It is of great interest to investigate how atypical maps effect derived quasar accretion disc constraints.
In general, observations are effected mainly by high magnification events.
For example, when convolving a magnification map with a given source profile, the result will depend more on the highly magnified pixels i.e. the high magnification tail of the MPD.
Upon constraining accretion disc models, part of the uncertainty of the results could be attributed to the uncertainty of the MPD.
As demonstrated, in certain parts of the $\kappa,\gamma$ parameter space the maps are almost statistically equivalent.
Therefore, for such $\kappa,\gamma$ values the MPDs are well-defined and an improvement on accretion disc constraints by using more than one map is not anticipated.
It would be of interest to examine to what extent an accretion disc model will differ when using MPDs that tend to fail the KS test with the corresponding $\langle\textrm{MPD}\rangle$, however, this is out of the scope of this paper.
In Table \ref{tab:2} we identify which of the macromodels of the known lensed systems shown in Fig. \ref{fig:existing} fall into regions of $\kappa,\gamma$ space where we have more than 10 per cent of failed pairs, or at least one MPD that failed the test with the $\langle\textrm{MPD}\rangle$ (for 24 $R_{\rm Ein}$ maps).
These are also the systems that will get more effected by microlens positions systematics if a map smaller than 24 $R_{\rm Ein}$ is used.
In conclusion, in some parts of the $\kappa,\gamma$ parameter space, using several maps with different microlens positions is expected to lead to more robust constraints in accretion disc models.
However, the magnitude of this improvement, and consequently the gain of following a many-map approach, is something to be determined in practice.
In any case, a map larger than 24 $R_{\rm Ein}$ most likely shows an average behaviour.

The results presented above constitute the most complete covering of the convergence and shear parameter space accomplished to date.
The different properties of the MPDs across the $\kappa$,$\gamma$ parameter space should be considered in designing a microlensing parameter survey, like the one proposed in BF12.
%$$$$$$$$$$$$$$$$$$$$$$$$$$$$$$$$$$$$$$$$$$$$$$$$$$$$$$$$$$$$$$$$$$$$$$$$$$$$$$$$$$$$$$$$$$$$$$$$$$$$$$$$$

%$$$$$$$$$$$$$$$$$$$$$$$$$$$$$$$$$$$$$$$$$$$$$$$$$$$$$$$$$$$$$$$$$$$$$$$$$$$$$$$$$$$$$$$$$$$$$$$$$$$$$$$$$
\section*{Acknowledgements}
The authors would like to thank Nick Bate and Darren Croton.
This work was performed on the gSTAR national facility at Swinburne University of Technology.
gSTAR is funded by Swinburne and the Australian Government's Education Investment Fund.
Our referee provided very helpful suggestions, which improved the quality of the wrok.
%$$$$$$$$$$$$$$$$$$$$$$$$$$$$$$$$$$$$$$$$$$$$$$$$$$$$$$$$$$$$$$$$$$$$$$$$$$$$$$$$$$$$$$$$$$$$$$$$$$$$$$$$$

\bibliographystyle{mn2e}
\bibliography{systematics}

%$$$$$$$$$$$$$$$$$$$$$$$$$$$$$$$$$$$$$$$$$$$$$$$$$$$$$$$$$$$$$$$$$$$$$$$$$$$$$$$$$$$$$$$$$$$$$$$$$$$$$$$$$
\appendix
\section{Map accuracy and the direct inverse ray-shooting technique}

\begin{figure}
\begin{center}
\includegraphics[width=0.47\textwidth]{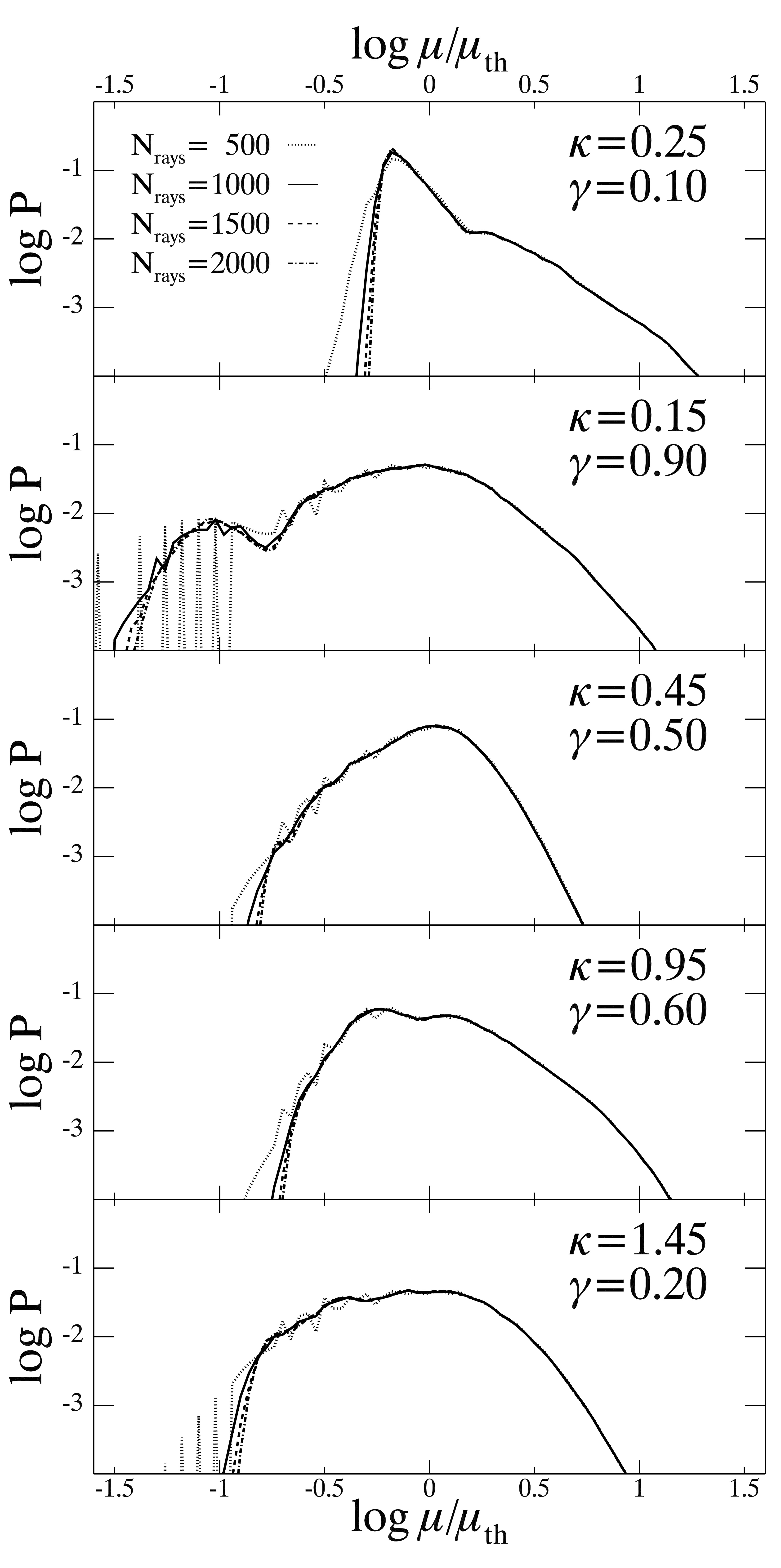}
\end{center}
\caption{Varying the accuracy in 5 representative regions of the $\kappa,\gamma$ parameter space. For each $\kappa,\gamma$ value we have calculated the same map (same microlens positions) using a different total number of rays, as indicated by the input parameter $N_{\rm rays}$. The units and the ranges of the figures are the same as in Fig. \ref{fig:single}.}
\label{fig:navg}
\end{figure}

\begin{figure}
\begin{center}
\includegraphics[width=0.47\textwidth]{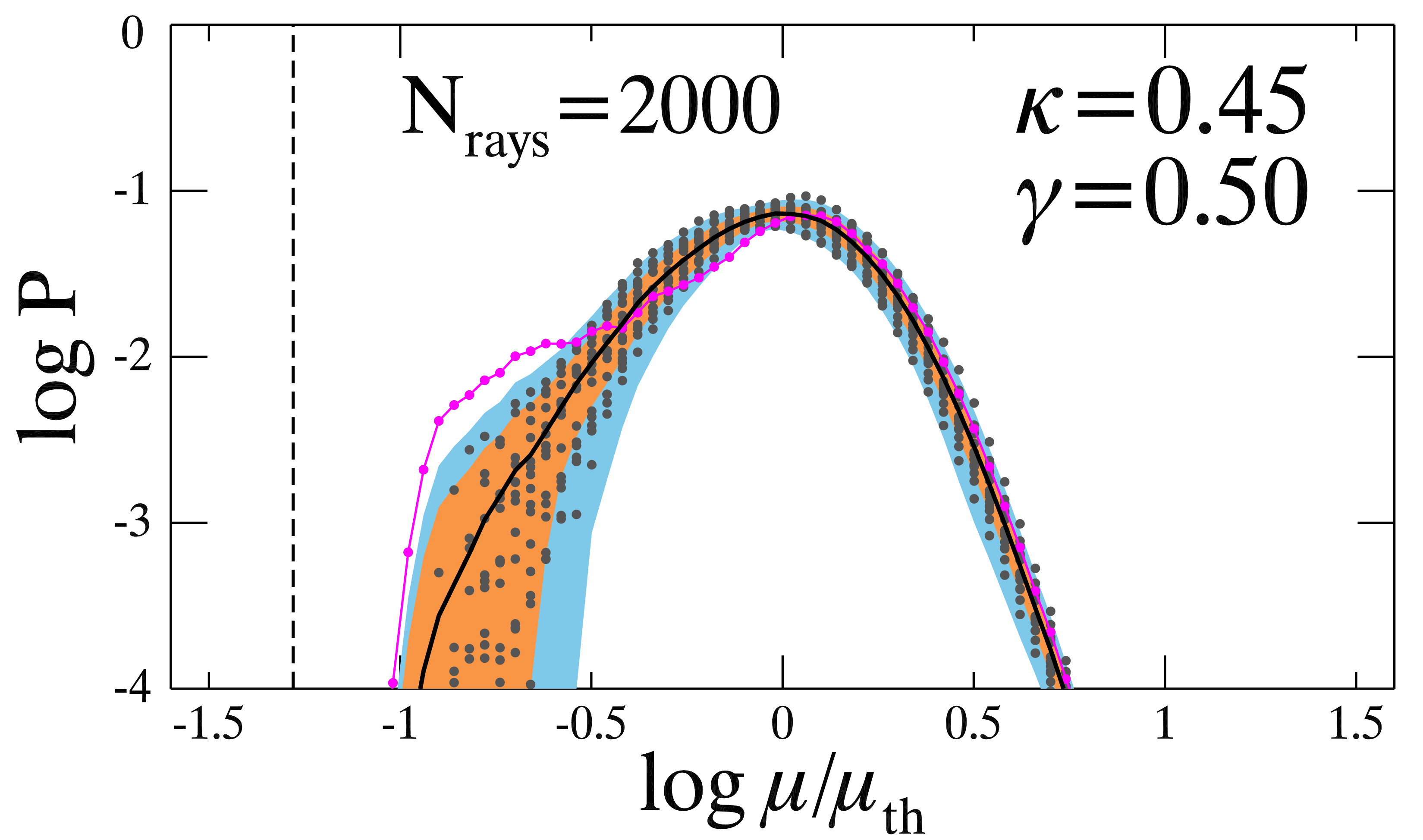}
\end{center}
\caption{We calculated the same 15 maps whose MPDs are shown in Fig. \ref{fig:single} using twice as many rays, $N_{\rm rays}=2000$. The minor effect of varying the microlens positions, as described in Section \ref{sec:discussion}, persists.}
\label{fig:single_navg}
\end{figure}

\begin{figure*}
\begin{center}
%\vspace{5cm}
\includegraphics[width=\textwidth]{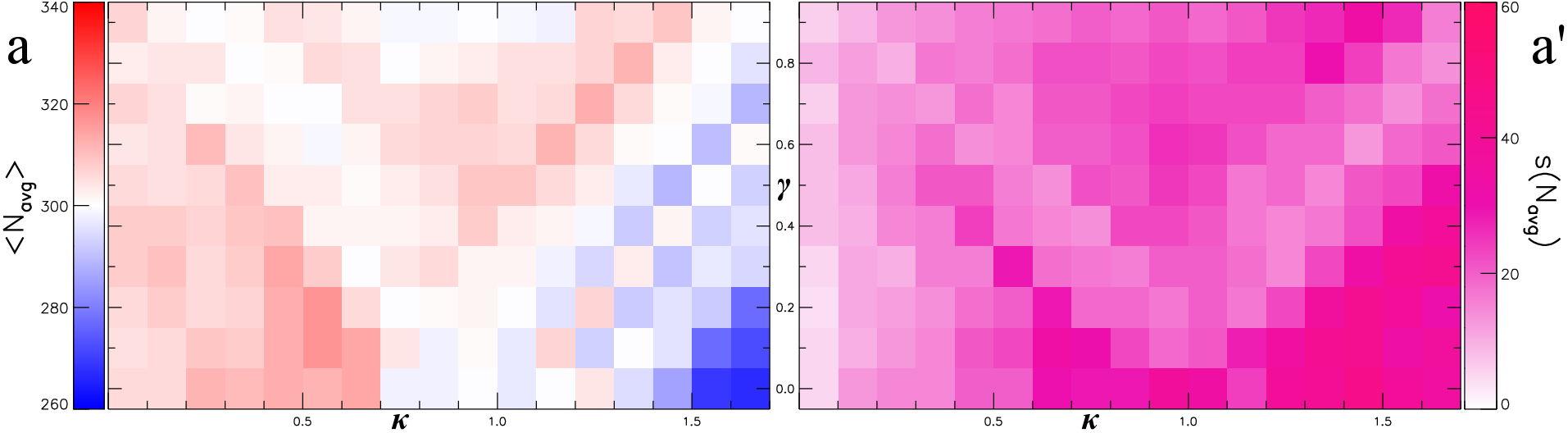}
\end{center}
\caption{We show the mean (panel a) and the standard deviation (panel a') of the average number of rays per map pixel, $N_{\rm avg}$, among 15 maps, across the $\kappa,\gamma$ parameter space. We compare the results with the value of 300 (white), which lies very close to the calculated mean $N_{\rm avg}$ among all the 2550 maps ($302\pm24$). Higher values are shown in red and lower in blue.}
\label{fig:stats_app}
\end{figure*}

Each pixel of a magnification map must have a number of rays large enough to accurately estimate the magnification of the enclosed area of the source plane.
In order to achieve this even for the low magnification pixels, we typically have to shoot billions of rays.
The total number of rays, $N_{\rm tot}$, which we are using for each map is determined through:
\begin{equation}
\label{eq:Ntot}
 N_{\rm tot} = N_{\rm in}^2 \times 10^4 \, ,
\end{equation}
where $N_{\rm in}$ is an input parameter, same as per equation 6 of \cite{Bate2010}.
For all our maps we have set $N_{\rm in}=1000$, leading to $N_{\rm tot}=10^{10}$.
In this appendix we demonstrate that this is a reasonable trade off between accuracy and computational time.

We have chosen 5 pairs of $\kappa,\gamma$ values across the parameter space, for which we calculate a magnification map with \textit{the same} microlens positions and $N_{\rm in}$ set to 500, 1000, 1500 and 2000.
The choice of $\kappa,\gamma$ values aims at representing distinct parameter space regions i.e. the minima, saddle-point, maxima and critical line regions.
The corresponding MPDs in each case are shown in Fig. \ref{fig:navg} (the axis ranges and units are the same as in Figs \ref{fig:single} and \ref{fig:distrs}).
We notice that changing the value of $N_{\rm in}$ has an effect only towards low magnifications ($<\mu_{\rm th}$).
For $N_{\rm in}=500$ (dotted line) the appearance of dips on the MPD for all $\kappa,\gamma$ values indicates that some magnification values are missing; map pixels corresponding to low-magnifications, which would have been reached by a small number of rays, were not reached by any because we did not use enough.
For $N_{\rm in}=1000$ (solid line), $1500$ (dashed line) and $2000$ (dotted-dashed line) the MPDs are essentially indistinguishable above $\mu_{\rm{th}}$.

Next, we examine the effect of changing the random microlens positions at increased map accuracy.
In Fig. \ref{fig:single_navg} we show the MPDs of 15 maps for $\kappa = 0.45$ and $\gamma = 0.5$, calculated with exactly the same random microlens positions as the 15 maps in Fig. \ref{fig:single}, but with twice as many rays per map (and twice the computational time).
We see that the microlens position effect persists and only a minor smoothing of the MPDs at low magnification occurs.
Because the computational time scales linearly with $N_{\rm{in}}$ \citep[equations 7 and 8 in][]{Thompson2010}, we find $N_{\rm in}=1000$ to be the best choice for our calculations.

Finally, we examine parameter space behaviour of $N_{\rm avg}$ (see Section \ref{sec:map-parameters}).
The calculated $N_{\rm avg}$ among all 2550 maps is $302\pm24$, however, in Fig. \ref{fig:stats_app} we show its behaviour across the $\kappa,\gamma$ parameter space.
We calculate the mean and the standard deviation of the average number of rays per pixel of each map, $N_{\rm avg}$, among the 15 maps for each $\kappa,\gamma$ value across the parameter space.
In Fig. \ref{fig:stats_app}(a) we compare the results with a value of 300 (white).
We see that Figs \ref{fig:stats_app}(a) and (a') are identical to Figs \ref{fig:stats}(a) and (a') where we show the mean magnification, $\bar{\mu}$, of the MPDs.
This should be anticipated due to a simple relation between $N_{\rm avg}$ and the mean magnification of a map:
\begin{equation}
\bar{\mu} = \sum_{ij} \frac{\mu_{ij}}{N_{\rm pix}}
= \sum_{ij} \frac{N_{ij}/N_{\rm rays}}{N_{\rm pix}}
= \frac{\bar{N}}{N_{\rm rays}}
= \frac{N_{\rm avg}}{N_{\rm rays}} \, ,
\end{equation}
where $N_{\rm pix}$ is the total number of pixels and we have used equation \ref{eq:mu} in the second step.
The systematic behaviour seen in distinct parts of the $\kappa,\gamma$ parameter space i.e. over-estimating $\bar{\mu}$ in the minima region and under-estimating it in the maxima with respect to $\mu_{\rm th}$, is because the ray-shooting area used in the GPU-D approach depends on the size and shape of the minimum ellipse \citep{Bate2010}, which in turn depends on the ratio $|1-\kappa+\gamma|/|1-\kappa-\gamma|$.

We give an estimate of the efficiency of the GPU-D approach, with respect to the total number of rays used, $N_{\rm tot}$, across the $\kappa,\gamma$ parameter space.
After we perform the ray-shooting and count the rays in each pixel, we are left with a number of rays for which the whole series of calculations was performed, but they were excluded from the final map, $N_{\rm ex}$, or as a fraction of the total number of rays, $\alpha N_{\rm tot}$.
We can write the average number of rays per pixel of the map as:
\begin{equation}
N_{\rm avg} = \bar{N}
= \sum_{ij} \frac{N_{ij}}{N_{\rm pix}}
= \frac{N_{\rm tot}-N_{\rm ex}}{N_{\rm pix}}
= \frac{(1-\alpha)N_{\rm tot}}{N_{\rm pix}} \, ,
\end{equation}
where we used the fact that the number of rays in a map is equal to $N_{\rm tot}-N_{\rm ex}$ 
Replacing $N_{\rm tot}$ from equation \ref{eq:Ntot} and using the calculated $N_{\rm avg}$, we get $\alpha=0.52\pm0.04$ or an efficiency of $48 \pm 4$ per cent, averaged over the parameter space.
If instead of the calculated global $N_{\rm avg}$ we use the individual values across the $\kappa,\gamma$ parameter space, we essentially end up with the same parameter space behaviour of $\alpha$ as the one shown in Fig. \ref{fig:stats_app} for $N_{\rm avg}$; the GPU-D code is more efficient in the minima than in the maxima region.

The low efficiency of the GPU-D code directly reflects the brute force approach followed.
While it is possible to make use of more sophisticated algorithms (tree code; \citealt{Wambsganss1990a}, inverse polygon mapping; \citealt{Mediavilla2006,Mediavilla2011b}) to increase the GPU-D efficiency, it may prove to be a quite complex task.
Instead, we argue that a brute force GPU approach in combination with a GPU cluster like gSTAR can lead more quickly to practical results.
%$$$$$$$$$$$$$$$$$$$$$$$$$$$$$$$$$$$$$$$$$$$$$$$$$$$$$$$$$$$$$$$$$$$$$$$$$$$$$$$$$$$$$$$$$$$$$$$$$$$$$$$$$

\label{lastpage}
\end{document}